\documentclass[showpacs, amsmath,amsfonts, prb, twocolumn]{revtex4}

\usepackage{amsmath}  % needed for \tfrac, \bmatrix, etc.
\usepackage{amsfonts} % needed for bold Greek, Fraktur, and blackboard bold
\usepackage{graphicx} % needed for figures

\begin{document}

\title{Qualitative methods in condensed matter physics}

\author{Maria Patmiou}
\email{maria.patmiou@rockets.utoledo.edu} % optional
\affiliation{Department of Physics and Astronomy, University of Toledo, Toledo, OH 43606, USA}
\author{V. G. Karpov}
\email{victor.karpov@utoledo.edu}
\affiliation{Department of Physics and Astronomy, University of Toledo, Toledo, OH 43606, USA}

\date{\today}
\begin{abstract}
Expanding on our former hypothesis that, in the current information age, teaching physics should become more intuition-based and aiming at pattern recognition skills, we present multiple examples of qualitative methods in condensed matter physics. They include the subjects of phonons, thermal and electronic properties of matter, electron-phonon interactions and some properties of semiconductors.
\end{abstract}
\maketitle

\tableofcontents

\section{Introduction}\label{sec:intro}
This manuscript expands on our recent premise \cite{karpov2019}
that teaching physics in the information age should become more intuitive and induction-driven rather than guided by memorization and principles assumed by deductive learning. The underlying reality is that the standard physics curriculum information is now available at our fingertips through smartphones, computers, etc. That information avalanche is accentuating the need for efficient pattern-recognition and result evaluation skills.

Our implied inductive learning (i. e. from examples) does not necessarily lead to generalizations (making the subsequent learning deductive \cite{howpeoplelearn}), which are neither sufficient nor necessary. Examples of inductive learning without generalization include language learning in children, sports training, professionally successful problem solving based on “experience”, etc.  In a similar manner, when learning the language of science -in this case physics- our brain’s functionality is geared toward induction through the processes of pattern recognition and neural network training.

A strong `experimental' support for the latter insight comes from the modern understanding of neural networks that demonstrates how such learning can be prevailing and useful. Indeed, studying the available machine learning and artificial intelligence operations does not offer a possibility to determine how the system makes its decision: all in all, it appears as a result of forming certain connections between the system neurons achieved by trial and error.  The resulting neural nets are created without a prior knowledge, neither can they be translated into operational theoretical deduction.

The pathway toward developing inductive learning abilities taken in our preceding work \cite{karpov2019} is that of practicing various qualitative methods. They are fast and intuitive, boosting the learner's self-esteem, benefiting both students and professionals in navigating the overwhelming amount of information. That preceding work was limited to qualitative methods in quantum mechanics, which is a well-established homogeneous field. The current manuscript aims at expanding over a different type of discipline, condensed matter physics, which is both multi-faceted (phonons, electrons, magnetism, elasticity, thermodynamics, etc.) and multi-format (condensed matter theory, material sciences, physics of semiconductors, electronic materials, etc.) \cite{yeh} Developing inductive learning abilities (pattern recognition, approximations, back of the envelope calculations, etc.) presents a formidable challenge. It is met here with the same approach as before:  demonstrating multiple patterns of qualitative analyses for various problems. By no means can we claim that this approach is sufficient; however, we hope it will be useful in developing such abilities in the field of condensed matter sciences.

But how can one make a representative selection of topics from the almost infinite number of condensed matter possibilities? Our approach here favors taste-based topics {\it allowing qualitative methods}. (We note parenthetically that many problems in condensed matter would not allow qualitative methods, such as e. g. energy analyses of particular microscopic configurations in solids. The ability to recognize `a smell' of qualitatively treatable problems appears to be a useful skill, and that can be instilled by examples.) Other contributions of condensed matter subjects attainable to qualitative methods are called upon for the benefit of the learners. The material that follows will remain an illustration.

One other goal here is to bring a flavor of condensed matter physics and sketch its skeleton for those without any significant background. We hope that such a sketch and intuition will allow faster progress by then consciously collecting and organizing relevant information from other sources.

We shall end this introduction admitting that the standard condensed matter curricula will remain utterly important in professional development. Our additions here will just facilitate their learning making the subject more intuitive and paving ways to inductive learning in other subfields of condensed matter physics.

\section{The ultimate simplicity models}\label{atomic}

It is obvious that a single atom cannot serve as a model of condensed matter for $N\gg 1$ atoms. However, two atoms do represent some basic properties of condensed matter when they form a diatomic molecule, since the latter possess a variety of relevant characteristics: binding energy, elasticity and anharmonicity, vibrational frequency, specific heat, etc. These characteristics are all practically significant determining the range of various material applications.

Another useful simplistic model in condensed matter physics is that of two level systems, which generally does not specify the physical nature of energy spectrum limited to the irreducible minimum of just two (a one level model is of no interest, not allowing any transitions in the system). Oftentimes, the two level system properties exhibit themselves in objects with much more complex energy spectra; in other cases, the two-level structure per se is dictated by the underlying physics.

The toy models below represent a useful example of simplifications where phenomena observed in complex systems (solids) are traced down to irreducibly primitive objects: two atoms or two energy levels. In a sense, they show how the number of two (but not one!) may be large enough to reveal the physics of many-body systems.

\subsection{Diatomic model}\label{sec:diat}

This simplistic model introduces a number of properties central in condensed matter physics: binding energies, adiabatic description, harmonic vibrations, elasticity and anharmonic effects, and deformation potential interaction.

\begin{figure}
\includegraphics[width=0.47\textwidth]{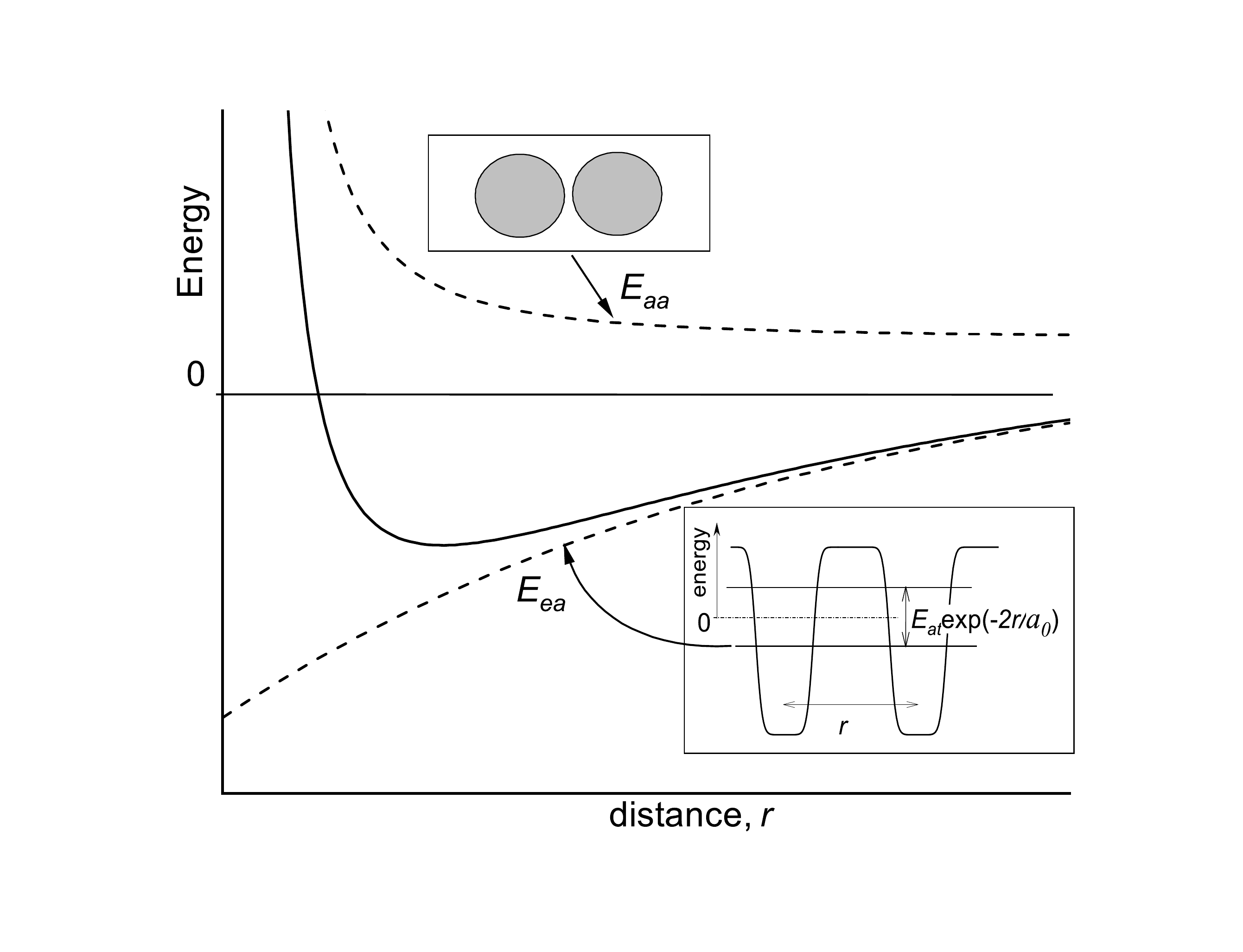}
\caption{A qualitative form of potential energy of two atoms as a sum of repulsive core-to-core interaction due to the valence band formation.  \label{Fig:diatorigin}}\end{figure}

We start with recalling the shape and origin of a diatomic interaction in Fig. \ref{Fig:diatorigin} where the potential energy is a sum of two contributions presented by dashed lines and illustrated in the insets. One of them is the repulsion between atomic cores $E_{aa}$ (for example between two protons in the case of hydrogen molecule). Another one, $E_{ea}$ is the attraction due to forming a valence bond with energy corresponding to lowest of two split levels originating from quantum tunneling of the valence electron between two atoms. We further consider that phenomenon in Sec. \ref{sec:locbas} below.

The latter energy is written in the adiabatic approximation \cite{ashcroft} where it is a function of atomic coordinates only but not their velocities. (We recall that the adiabatic approximation calculates the electron energy at a given instantaneous atomic configuration, neglecting atomic motion as very slow due to their heavy masses compared to that of electrons.) The corresponding atomic energy is calculated as the average over the electron wave function corresponding to a given set of atomic positions. The small parameter describing the slowness of atomic systems compared to that of electrons is related to the ratio of the electronic and atomic masses as demonstrated in Eq. (\ref{eq:spring}) below.

Note that $E_{ea}$ originally represents the electronic energy vs. interatomic distance. Its time-average over the fast electronic movements (according to the adiabatic principle), is the potential energy of atomic nuclei. However, its electronic origin becomes important when we address the question of how the electronic energy will respond to changes in interatomic distance. As is seen in Fig. \ref{Fig:diatorigin}, $E_{ea}$ decreases when that distance becomes smaller, i. e. it is energetically favorable for the electrons to shorten a molecule. For example, adding one more electron with the same energy $E_{ea}$ will shift the balance (with $E_{aa}\rightarrow 2E_{aa}$) towards shorter lengths, thus shrinking the molecule. Such interactions between the electrons and atoms are commonly referred to as the deformation potential.

We mention here just one example of deformation potential responsible for sound absorption in metals. As is well known, a sound wave produces periodically distributed regions of higher and lower material density, which, through the deformation potential, trigger the electron redistribution towards denser regions where they have lower energies. The electric currents corresponding to such a redistribution will result in energy dissipation proportional to the metal resistivity; hence, the original energy of a sound wave transforming into Joule heat, which is sound absorption. We will go through a number of other examples of deformation potential interaction in Sec. \ref{sec:e-ph} below.

\begin{figure}[bt]
\includegraphics[width=0.47\textwidth]{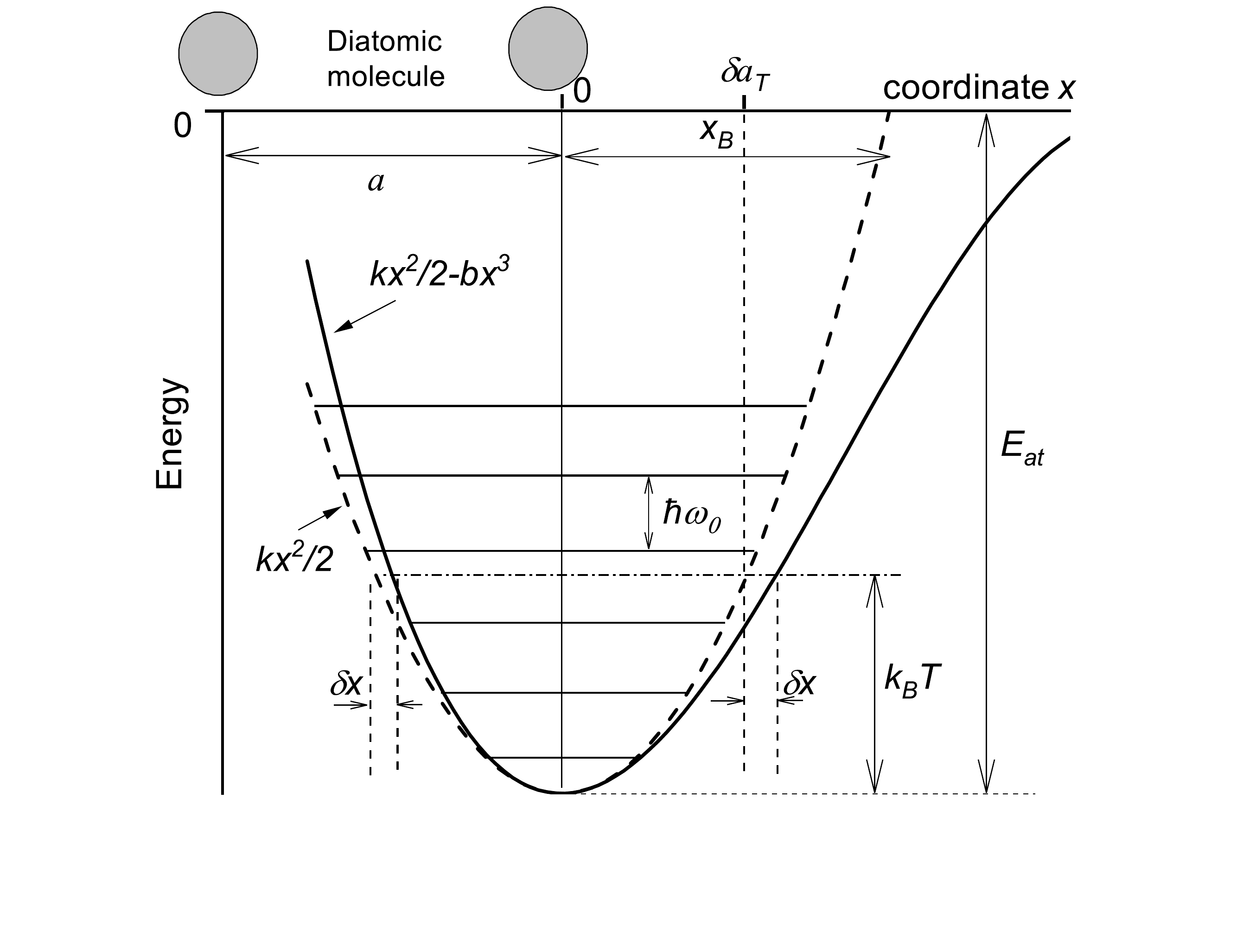}
\caption{A sketch of potential energy and energy levels (not to scale) in a diatomic molecule presented with two circles in the top left corner. The dashed curve stands for the harmonic approximation. The horizontal domains represent the energy levels of a harmonic oscillator (separated by gaps of $\hbar \omega _0$), for definiteness chosen here to be significantly smaller than the characteristic thermal energy $k_BT$. $a$ is the equilibrium length of a molecule, $x_B$  is the elongation that would break it increasing $kx^2/2$ above the binding energy $E_{at}$ (note that the leftmost 0 in the figure corresponds to the energy axis). $\delta a_T$ is the amplitude of harmonic thermal vibrations, $\delta x$ is the change in that amplitude due to anharmonism.  \label{Fig:diatomic}}\end{figure}

In integral, our diatomic model is illustrated in Fig. \ref{Fig:diatomic}. The characteristic linear and energy scales are given by $a_0\sim 2-3$ $\AA$ and $E_{at}\sim 3-5$ eV, as dictated by atomic physics. In what follows we accept for simplicity the rough order of magnitude estimates $a\sim 1$ $\AA$ and $E_{at}\sim 10$ eV. They are interrelated through $E_{at}\sim \hbar ^2/(m_ea^2)$, which interprets $E_{at}$ as the characteristic energy of the molecule binding electron localized in a linear domain of $a$.

\subsubsection{Harmonic approximation} In the harmonic approximation, the potential energy of the molecule is $kx^2/2$ and the harmonic oscillation frequency is $\omega _0=\sqrt{k/M}$ where $M$ is the atomic mass (strictly speaking, the reduced atomic mass of the molecule, but that subtlety is not important here), which for the order of magnitude estimates we will take as $M\sim 10^{-22}$ g corresponding to the middle of the periodic table of elements. In harmonic approximation, the molecule breakup takes place at the characteristic stretch $x_B$ when $kx_B^2/2\sim E_{at}$ (see Fig. \ref{Fig:diatomic}). Because we do not have any characteristic length at hand other than the atomic length $a$, we expect $x_B\sim a_0$, which yields, $k\sim E_{at}/a_0^2$, resulting in
\begin{equation}\label{omega0}
\omega _0\sim 3\times 10^{13}\quad {\rm s}^{-1}\quad {\rm and}\quad \hbar\omega _0\sim 0.03 \quad {\rm eV},\end{equation}
well within the ballpark of the accepted values for the frequencies and energies of atomic vibrations for both the molecules and condensed matter.

Note that the vibrational energy can be represented in the form,
\begin{equation}
\hbar\omega _0\sim E_{at}\sqrt{\frac{m_e}{M}}
\label{eq:spring}\end{equation} where $m_e/M\ll 1$ ($m_e/M\sim 10^{-5}$ with parameters here) is the famous adiabatic
ratio.

Furthermore, the characteristic vibration amplitude at zero temperature can
be estimated as the De Broglie wavelength,
\begin{eqnarray}\delta a_0\sim
\frac{\hbar}{p}\sim\frac{\hbar}{\sqrt{M\hbar\omega_0}}\sim
a\left(\frac{m_e}{M}\right)^{1/4},\label{eq:ampl}\end{eqnarray}
of the order of several percent of the characteristic
atomic length.

In connection with the latter, we note the high temperature vibration amplitude $\delta a _T$
that increases with temperature $T$ (Fig. \ref{Fig:diatomic}). For high enough temperatures  $k_BT\gg \hbar\omega _0$  (where $k_B$ is the Boltzmann's constant), the turning points condition yields,
$k \delta a_T^2\sim k_B T$, and
\begin{equation}\label{eq:deltaT}\delta a_T\sim\sqrt{\frac{k_BT}{k}}\sim\sqrt{\frac{k_BT}{\hbar\omega_0}}\delta a_0.\end{equation}
When $\delta a _T$ increases above a certain fraction (typically 10 percent)
of $a$, the material melts. This constitutes the well known
(empirical) Lindeman criterion of melting. For our purposes, that criterion tells that the atomic vibration amplitudes are relatively small compared to the interatomic distances.

In the harmonic approximation, the diatomic model predicts that the internal energy equals the sum of averages of potential and kinetic energies, which are both equal to $k_BT/2$ in the classical limit $k_BT\gg \hbar\omega _0$ (equipartition theorem). The derivative of that energy versus $T$ gives the specific heat, $c=k_B$. For the low temperature quantum limit, $k_BT\ll \hbar\omega _0$, the probability of excitations from the ground level $\exp(-\hbar\omega _0/k_BT)$ becomes exponentially small, hence $c\rightarrow 0$. These predictions remain in the limiting cases of the {\it Einstein solid} model, that postulates equivalent harmonic vibration frequencies $\omega _0$ for all $N\gg 1$ atoms in a solid, predicting the heat capacitance
\begin{equation}\label{eq:einstein}
c=3Nk_B\left(\frac{\hbar\omega _0}{2k_BT}\right)^2\left[\sinh\left(\frac{\hbar\omega _0}{2k_BT}\right)\right]^{-2}\end{equation}
where the multiplier 3 reflects the fact that each atom can vibrate along three independent coordinates $x,y,z$. The exponentially small $c\propto \exp(-\hbar\omega _0/k_BT)$ predicted by Eq. (\ref{eq:einstein}) does not agree with the data, which reflects the inadequacy of the diatomic molecule model as explained below in Sec. \ref{sec:avs}.

\subsubsection{Anharmonic effects: thermal expansion} Based on the smallness of atomic vibration amplitudes, the potential energy of a diatomic molecule can be expanded beyond the harmonic approximation,
\begin{equation}\label{eq:anharm} U(x)=kx^2/2-\beta x^3+... \quad {\rm with} \quad k\sim E_{at}/a^2,\ \beta\sim E_{at}/a^3\end{equation}
where we have employed the estimate for $\beta$ combining the available parameters of needed dimensionality, along the lines of the above estimate for $k$. The negative sign in front of $\beta$ is taken on empirical grounds.

The difference between the harmonic and anharmonic approximation is illustrated in Fig. \ref{Fig:diatomic}. It demonstrates the increase of the right and decrease of the left turning point coordinates determined by the condition $U(x_T)=k_BT$; both changes are represented as $\delta x$ in $\delta a _T$ and a slight difference between them is neglected. We note now that in the harmonic approximation, the molecule vibrations are symmetric with respect to $x=0$ while the anharmonism makes the positive deformations larger in absolute values than the negative ones. The latter observation means that the average length of the molecule increases with temperature, thus representing thermal expansion.

The thermal expansion effect can be described more quantitatively by setting $U(\delta a_T+\delta x)=kT$ with $k\delta a_T^2/2=k_BT$ and $\delta x\ll \delta a_T$; here $\delta a_T$ is the high temperature amplitude of harmonic vibrations determined in Eq. (\ref{eq:deltaT}). This yields the thermal expansion linear in temperature, $\delta x=2\beta k_BT/k^2$ with the thermal expansion coefficient
\begin{equation}\alpha \equiv \frac{1}{a}\frac{d\delta x}{dT}=\frac{2k_B\beta}{k^2a}.\end{equation}

The thermal expansion is often described in terms of the dimensionless Gr\"{u}neisen parameter $\Gamma =\alpha K/c$ where $K$ is the bulk modulus and $c$ is the thermal capacity. By definition, $2K$ is the proportionality coefficient between the square of dimensional dilation, $(\delta a_T/a)^2$ and the elastic deformation energy, i. e. $K\sim ka^2$. Taking into account the heat capacity  $k_B$ and adopting the estimates for $k$ and $\beta$ from Eq. (\ref{eq:anharm}), one gets,
\begin{equation}
\Gamma\sim a\beta /k\sim 1,\end{equation}
in agreement with the data for high temperature Gr\"{u}neisen parameters of the majority of solids.

We will come across other links with the concept of diatomic molecules later in the text, in Sections \ref{sec:locbas}, \ref{sec:e-ph}, \ref{sec:etren}.

\subsection{Two-level systems}\label{sec:TLS}
\begin{figure}[bt]
\includegraphics[width=0.5\textwidth]{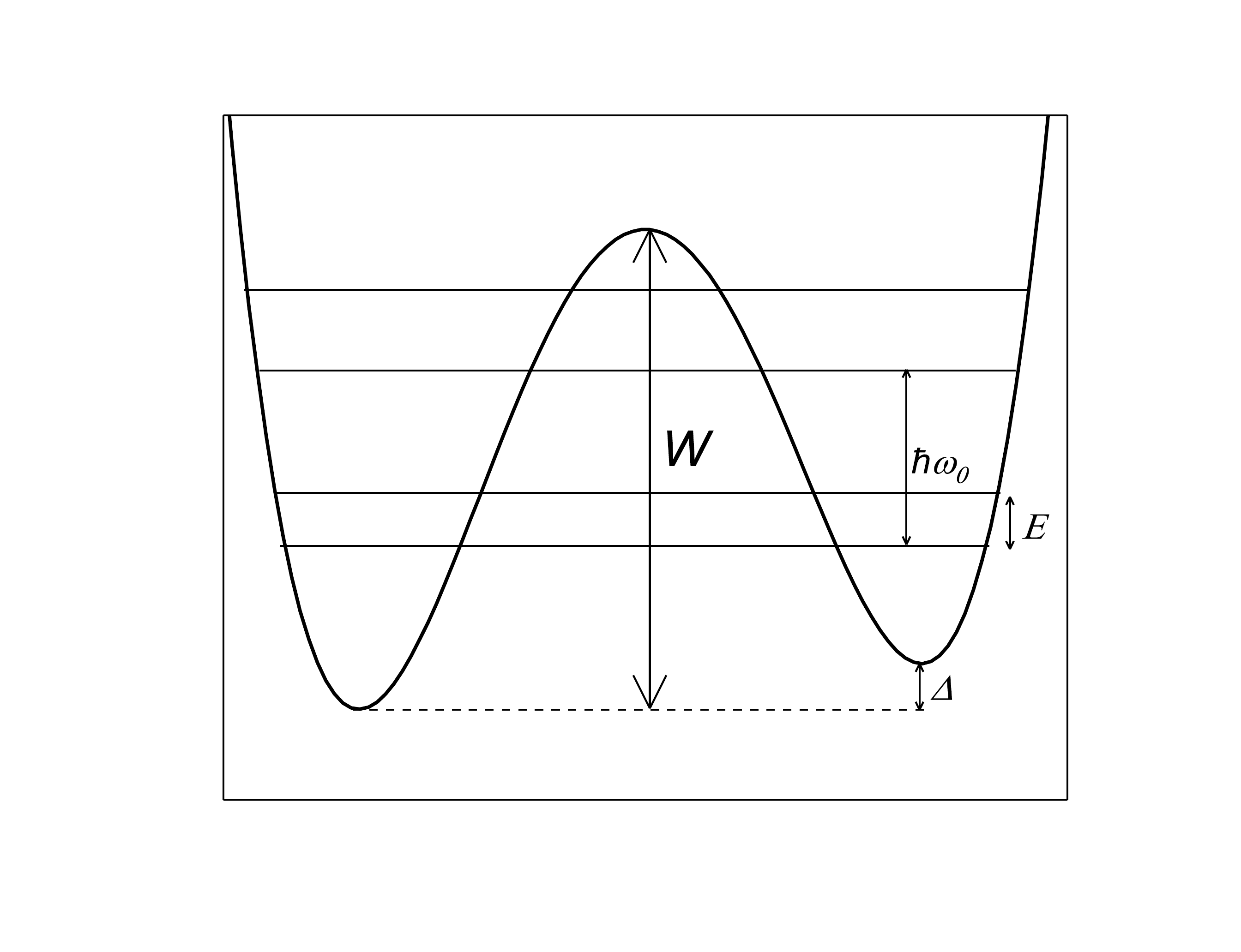}
\caption{Sketch of a double well atomic potential with energy levels corresponding to two classical wells with energy minima difference $\Delta$ and the lowest energy level gap $E$; not to scale: should be $\Delta <E\ll \hbar\omega _0\ll W$. \label{Fig:DWP}}\end{figure}

Consider a system with two energy levels, 0 and $E$. They can represent the electron spin energies in magnetic field $H$ (in which case $E=2\mu H$ where $\mu$ is the Bohr magneton) or the energies of certain nonequivalent configurations (say, off-centers) of ions in some crystals, or other atomic configurations possessing multiple equilibrium points in amorphous solids or liquids. For specificity, we will assume here atomic-type excitations. A model of two local minima (double well potential) along some unspecified configurational coordinate is presented in Fig. \ref{Fig:DWP}. That particular model diagram presents two-level excitations with energies $E$ considerably lower than the intra-gap energy levels $\hbar\omega _0$, which is achieved due to low asymmetry $\Delta$ of two well minima and low tunneling splitting $V_0$.

It is interesting that the double well potential model often gives a good description, even for systems that have more than two local minima. That happens when the transitions between various pairs of wells (states) are practically independent or when efficient transitions take place within only one such pair. The latter may be a consequence of a relatively low height $W$ of their separating barrier. Apparently, such a situation takes place in amorphous systems where local atomic configurations are random and their significant differences make only one of them efficient. Indeed, the double well atomic potential has been established as one of the central concepts for glasses. The underlying physics is that glasses are obtained by freezing of a material from its melted state. A rapid freezing leaves some microscopic volumes of liquid material arrested by momentarily frozen surroundings. In those arrested micro-volumes, some atoms or groups of atoms retain their liquid-like motility, being able to move between several potential minima. Of all possible potential barriers, which are significantly different due to the system randomness, one is the lowest and its related inter-well transitions dominate; hence, double well potentials.

A well known property of two-level systems is their contributions to the heat capacity $c(T)$ in the form of a peak at $T\approx E/k_B$ called the Schottky anomaly. Qualitatively, it can be understood by noting that for low $T\ll E/k_B$ the system cannot absorb any significant heat because there is not enough energy $kT$ to trigger upward transitions. In the opposite limit of $T\gg E/k_B$, both upper and lower energy levels are almost equally populated and practically no additional excitations are possible. Both the latter prohibitive factors are relaxed when $T\sim E/k_B$, where the specific heat must be at a maximum.

The same conclusion can be obtained more formally. For $N$ double well potentials, we use the definition $c=Nd\langle E\rangle /dT$ where $\langle E\rangle$ is the average energy per system. Using the standard definition of averages, $\langle E\rangle =0\times p_0 +E\times p_E=E\times p_E$ where $p_0$ and $p_E$ are the probabilities of finding a system in a state with energy 0 and $E$ respectively. The Gibbs distribution of thermodynamics dictates that $p_E=p_0\exp (-E/k_BT)$. Substituting the latter in the condition of unit total probability, $p_0+p_E=1$, yields $p_E=[1+\exp (E/k_BT)]^{-1}$. Differentiating $\langle E\rangle =Ep_E$ gives the Schottky contribution to the heat capacity,
\begin{equation}\label{eq:Sch}c_S=\frac{Nk_B\exp (E/k_BT)}{[1+\exp (E/k_BT)]^2}.\end{equation}
As a function of $T$,- it represents the Schottky anomaly, a peak in the vicinity of $T=E/k_B$, whose feature has been observed.

Furthermore, the double well potential model can include interaction between the two states by allowing quantum tunneling and its corresponding splitting $V_0$. Taking such splitting into account results in two states collectivized between the wells. The energy gap between them is given by,
\begin{equation}\label{eq:DWPE}E=\sqrt{\Delta ^2 +V_0^2}.\end{equation}
Eq. (\ref{eq:DWPE}) follows from the perturbation theory for adjacent levels described in the Sec. \ref{sec:app}. The model of Eq. (\ref{eq:DWPE}) with random $\Delta$ and $V_0$ describes a remarkably broad variety of phenomena in multiple glasses, such as specific heat, thermal conductivity, sound absorption, thermal expansion, etc. \cite{galperin1989}

As a simple example, consider briefly the specific heat in glasses. Due to their inherent randomness, we accept that the energy gaps $E$ are random in the ensemble of double well potentials. If we concentrate on the low temperature properties with $k_BT\ll \hbar\omega_0$, then excitations with $E\ll \hbar\omega_0$ will dominate the observations. Because $E$ is much below any characteristic energies in a solid ($\hbar\omega _0$, $E_{at}$, etc.) there are no energy scales available to characterize their density of state $n(E)$ (DOS, number of states per energy pert volume) and we have to accept $n(E)=const$ for $E\ll \hbar\omega _0$. Therefore, the number of contributing states at temperature $T$ is estimated as $nk_BT$ and their related energy is $\langle E\rangle\sim kT\times (nk_BT)$. Differentiating the latter with respect to $T$ gives the low-temperature specific heat $c\sim nk_B^2T$. Of course, the same result could be obtained by integrating Eq. (\ref{eq:Sch}) over energies $E$. Such a specific heat -linear in $T$- was observed in a great number of glasses.

\begin{figure}[htb]
\includegraphics[width=0.37\textwidth]{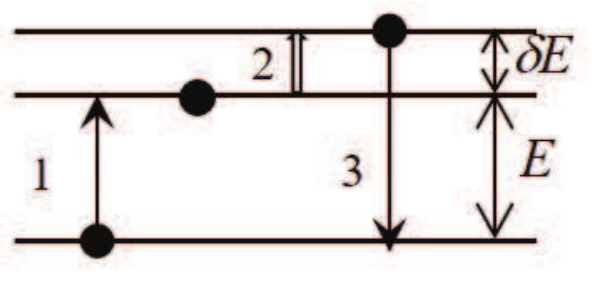}
\caption{Mechanism of the relaxation absorption by a two level system. The energy gap $E$ is modulated by the external field with amplitude $\delta E\ll k_BT$ (fat arrow 2). The upward thermal transition of a particle takes place when the energy of the two-level system is close to its minimum value while the downward transition (3) occurs close to its maximum (3); hence the energy difference $\delta E$ dissipates into heat which is tantamount to absorption. \label{Fig:absorption}}\end{figure}

Our second example describes EM field absorption due to two-level systems illustrated in Fig. \ref{Fig:absorption}. In the absence of external perturbations, the fluctuations in the difference in populations of the bottom and top levels $\Delta n$ is described by the standard relaxation equation $d\Delta n/dt=-\Delta n/\tau$ where $\tau$ is the relaxation time. That equation guarantees that fluctuations decay with time exponentially, $\Delta n\propto\exp(-t/\tau)$. The absorption is due to the time delay between the events of particle excitation and relaxation as shown in Fig. \ref{Fig:absorption}. It is clear that when the frequency of oscillations is much higher than $1/\tau $, then very fast oscillations average out the absorption process. On the other hand, very slow oscillations allow the energy level populations to adiabatically follow their instantaneous positions; hence, suppressed dissipation. We arrive at the conclusion that dissipation must be a maximum at frequencies of the order of $1/\tau$.

More formally, the external perturbation introduces the energy modulation with amplitude $\delta E\ll k_BT$ and frequency $\omega$ which translates into a stimulated modulation of the population difference $\delta E/k_BT$. The corresponding relaxation equation becomes $$d\Delta n/dt=-\Delta n/\tau +(\delta E/k_BT)\exp(i\omega t).$$ Its solution takes the form of $\Delta n=\alpha \delta E \exp(i\omega t)$ with the susceptibility $\alpha =1/(1+i\omega t)$. The absorption coefficient is proportional to the imaginary part of the latter, i. e. $I\propto \omega\tau /[1+(\omega\tau )^2]$, a maximum at $\omega\tau \sim 1$. (More in detail derivation of the same is given by Kittel. \cite{kittelthin})

\section{Delocalized atomic vibrations: phonons and fractons}\label{sec:avs}

The atomic vibrations described in Sec. \ref{sec:diat} are confined to certain local entities such as diatomic molecules or double well atomic potentials. In condensed matter, each atom interacts with many others, all  of  them contributing to vibrations in a rather non-additive manner. In particular, we will see how the vibrational spectrum of an $N$-atomic linear chain is very different from the superposition of spectra of its constituting pairs of atoms.

\begin{figure}[bt]
\includegraphics[width=0.4\textwidth]{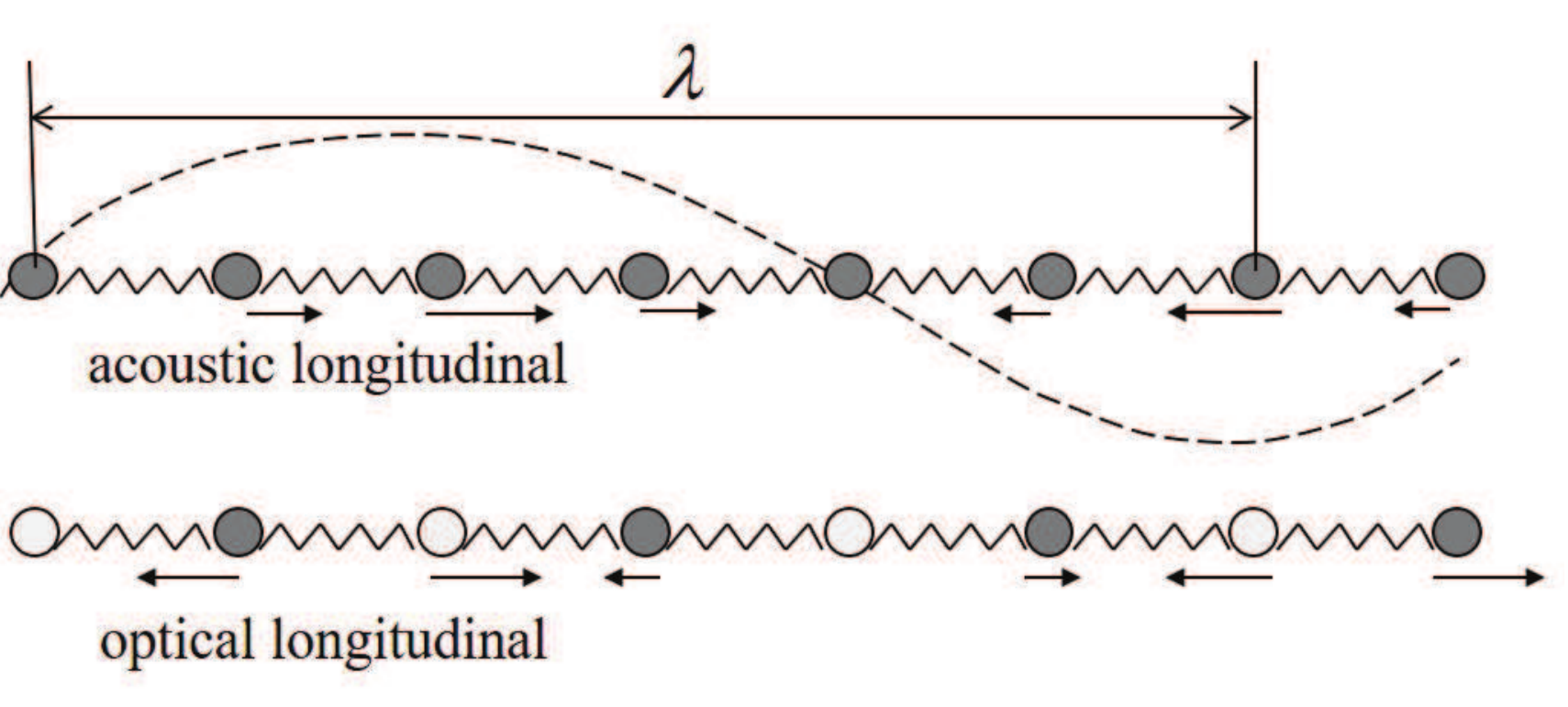}
\caption{Atomic displacements for acoustic (top) and optical
(bottom) vibrations \label{Fig:displ}}\end{figure}

More specifically, interactions of multiple coupled atoms make their vibrational excitations delocalized and spreading over the entire chain. Such excitations have a structure of waves, and quantization of such waves leads to the concept of phonons. Just like photons are the quanta of electromagnetic energy, phonons are the quanta of vibrational waves. The delocalized atomic vibrations in condensed matter exhibit two major modes: acoustic and optical (Fig. \ref{Fig:displ}). In what follows, we neglect such additional features of acoustic and optical waves as their possible longitudinal or transverse nature displacements, as they do not have any significant effect on their qualitative properties.

We will start by considering the acoustic mode. In the acoustic mode the particles move coherently around their equilibrium positions such that there are places where the atoms are closer together and others where they are farther apart. (Fig. \ref{Fig:displ} (top) describes such squeeze - stretch motion for a system of identical atoms). This dynamic is similar to that of propagation of sound, which is the reason it is called acoustic. The angular frequency $\omega$ can be described by the standard concept of harmonic vibrations of a certain mass $M_{eff}$ attached to a spring of elasticity $k_eff$:
%The consideration will be  based on a simple expression for the
%frequency of elastic spring vibrations,
\begin{equation}
\omega =\sqrt{\frac{k_{eff}}{M_{eff}}}.\label{eq:spring1}\end{equation} where
here, the effective spring constant mass will depend on the vibration wavelength.

Consider elastic waves of wavelength $\lambda$ (and wavenumber $q$) in a linear chain with interatomic distance $a$ as in Fig. \ref{Fig:displ} (top),
\begin{equation}\lambda = Na \quad {\rm and}
\quad q\sim1/\lambda .\label{eq:lambda}\end{equation}
Here $N$ is the number of atoms per wavelength. A $\lambda$-domain dilation $\delta x$
translates into
$\delta x/N$ per individual interatomic bond; hence the force
and effective spring constant are smaller by the factor of $N$,
$$k_{eff}=\frac{k}{N}.$$ The effective mass of atoms taking part in such
vibrations is
$$M_{eff}=MN.$$

\begin{figure}[bt]
\includegraphics[width=0.52\textwidth]{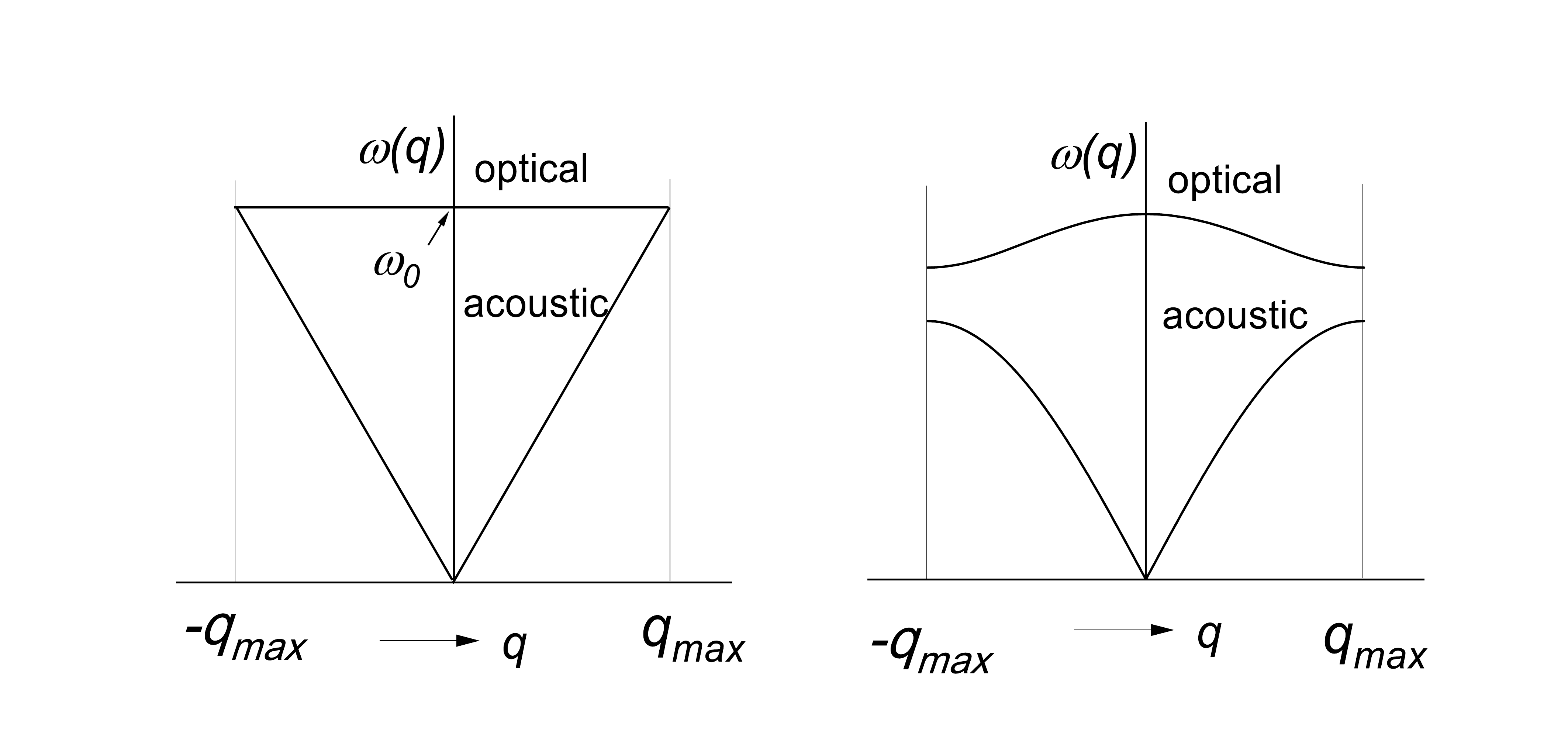}
\caption{Qualitatively predicted (left) and realistic (right)
phonon dispersion curves. \label{Fig:disp}}\end{figure}

Substituting these estimates into Eq. (\ref{eq:spring1}) and
taking into account Eq. (\ref{eq:lambda}) yields the dispersion
law (see Fig. \ref{Fig:disp})
\begin{equation} \omega = vq \quad {\rm with}\quad v=a\omega
_0 \label{eq:disp1}\quad {\rm and}\quad \omega
_0 =\sqrt{k/M}\end{equation} Here $v$ has the meaning of
sound velocity. Using the numerical values from Sec. \ref{sec:diat} we estimate $v\sim
3\cdot 10^5$ cm/s, in the ballpark of data for a variety of
solids. Also, we note that the minimum wavelength should be set
equal to the interatomic distance, $\lambda _{min}=a$. This corresponds to,
$q\leq q_{max}\sim 1/a$ in Eq. (\ref{eq:disp1}) (Note that we neglect numerical coefficients such as $2\pi$ in $q=2\pi /\lambda$.)

In the case of optical vibrations, the non-identical nearest neighbor atoms move in
pairs oscillating about their center of mass (bottom chain in figure  (Fig. \ref{Fig:displ}).
These vibrations with non-identical atoms bearing non-equivalent electric  charges form oscillating electric dipoles capable of interacting with electromagnetic fields; hence, the name of optical phonons; they
are responsible for light scattering and other optical
phenomena in solids. For optical vibrations, one
gets
$$k_{eff}=Nk,$$ because the deformation of $N$ springs will
take place simultaneously. The mass remains the same as in the previous
case, $M_{eff}=MN$. Substituting this again into Eq.
(\ref{eq:spring1}) gives,
\begin{equation} \omega=\omega
_0=const(q).\label{eq:disp2}\end{equation}

The above presented phonon dispersion laws are in qualitative
agreement with the measured curves (Fig. \ref{Fig:disp}). The
deviations in the region of $\lambda \sim a$ are due to strong
interference between the propagating waves and their reflections from atoms. Such reflections create
standing waves whose zero group velocities, $d\omega /dq=0$ correspond to flattening of the dispersion curves at $q\rightarrow
q_{max}$.

Looking at the same from a different angle, \cite{migdal} we note that since the system does not dissipate
energy, it must remain invariant under time reversal
($t\rightarrow -t$). Therefore, the equation
determining its frequency spectra can contain only even powers
of $\omega$. [For example, the Fourier transform of Newton's law for small vibrations,
$M\ddot{u}_n=-\sum _mF_{mn}u_m$ becomes $M\omega ^2 \tilde{u_n}=\sum
_mF_{mn}\tilde{u_m}$.] Consequently, it is $\omega ^2$ (and not
just $\omega$) that must be an analytical function of the
system parameters. In particular, for small wave-vectors,
$$\omega ^2=a+v^2q^2+bq^4+...$$
For sound waves, the limit $q\rightarrow 0$ corresponds to the
displacement of a system as a whole that does not change its
energy; hence, $a=0$ and $\omega =vq$. For optical vibrations,
there is no reason to nullify $a$; hence, $\omega
=\sqrt{a}=const$ when $q\rightarrow 0$.

\subsection{Phonon density of states}

Similar to photons, phonons exhibit both particle and wave properties. Each phonon is characterized by its
energy $E=\hbar\omega$ and momentum $p=\hbar q$ related through
the dispersion law $\omega (q)$. The phonon density of states
(DOS) $D(\omega )$ is defined as a number of phonon states per
unit frequency interval in the proximity of a given frequency,
$D(\omega )=dN (\omega )/d\omega$. The concept of phonon DOS plays an important role with multiple applications ranging from the thermodynamic to interactions of atomic vibrations with radiation and electrons.

For acoustic phonons, the number of states with
frequencies below $\omega$ can be counted by noting that they
are the same as the states with momenta below $p(\omega )$.
The corresponding phase volume, and thus the number of states
is proportional to $p^3$, the volume of a sphere of radius $p$
in the momentum space. Because $\omega\propto p$, (Eq. (\ref{eq:disp1}))this gives
$N(\omega )\propto \omega ^3$ and $D(\omega )\propto \omega
^2$.

The proportionality coefficient in the latter relation will
depend on the volume of the system and can be readily estimated from its
dimensionality. For example, DOS per atom will have the
dimension $\omega ^{-1}$. Having only the characteristic
frequency ($\omega _0$) at hand, the choice becomes:
$$D(\omega )\sim\frac{\omega ^2}{\omega _0^3}.$$ This formula
can be easily generalized to the case of arbitrary dimensions
$d=1,2,3,..$,
\begin{equation}D(\omega )\sim \frac{\omega ^{d-1}}{\omega _0
^d}.\label{eq:phononDOS}\end{equation}

Strictly speaking, the coefficient in the latter relations
should be corrected to preserve the number of degrees of
freedom per atom, which was originally 3 (before the atom is made
a part of a solid). Therefore we require $$\int D(\omega )d\omega =3.$$ To
maintain the latter relationship, the upper limit in the
integral is chosen to be somewhat different from $\omega _0$.
Namely, it should be defined as the {\it Debye} frequency,
\begin{equation}
\omega _D=(3d)^{1/d}\omega _0.\label{eq:Debye}\end{equation} As
a result
\begin{equation}D(\omega )= \frac{\omega ^{d-1}}{\omega _D
^d}, \quad \int _0^{\omega _D}D(\omega )d\omega =3.
\label{eq:phononDOS1}\end{equation} In most practical
applications, however, the difference between $\omega _D$ and
$\omega _0$ is insignificant.

For the case of optical phonons, the dispersion is not very
significant, DOS is large in a limited energy interval and can
be approximated by the delta-function,
$$D(\omega )\sim\delta (\omega - \omega _0).$$
The proportionality coefficient depends on the microscopic
structure of a primitive cell that may have several atoms
giving rise to different types of phonon polarization.

\subsection{Specific heat}

By definition, the specific heat of a system is given by:
$$C=\frac{dU}{dT}$$ where $U$ is the system energy. Phonons are traditionally treated with Bose-Einstein statistics where the number of phonons $dN=n\hbar d\omega$ with energy in the interval $\hbar d\omega$ has is described through the occupation number
$n(\hbar\omega)=1/(\exp(\hbar \omega/k_BT)-1)$. In particular, the energy $U$ can be evaluated by integration of the product $n(\hbar \omega )D(\omega )$ over all phonon frequencies.

We now present a simpler argument for the evaluation of phonon specific heat.
At low
temperatures, such that $k_BT\ll \hbar\omega _0$, one gets $n(\hbar\omega_0)\sim \exp (-(\hbar\omega_0/k_BT))\ll 1$. Therefore the
probability of optical excitations and their contributions to
$U$ are negligibly small,
%exponentially small ($\exp (\hbar\omega _0/k_BT)\ll
%1.$)
and we can take into account only acoustic phonons. Their
energy is estimated as the number of active phonon states
$N$ with energies $\hbar\omega _0\lesssim k_BT$ times the
thermal energy $k_BT$,
\begin{equation}U\sim kTN(\omega =k_BT/\hbar )\sim
k_BT\frac{(kT)^d}{(\hbar\omega _0)^d}.\label{eq:Upot}\end{equation} Therefore,
\begin{equation}C\sim k_B\left(\frac{T}{T_D}\right)^d\quad {\rm
where}\quad T_D=\frac{\hbar\omega
_D}{k_B}\label{eq:DebyeT3}\end{equation} and $T_D$ is known as
the Debye temperature.\cite{ashcroft}

As the temperature increases towards $T_D$, all the available
states become affected, their number saturates, and so does the
specific heat.

\subsection{Thermal expansion and Gr\"{u}neisen parameter}

So far we used the harmonic oscillator (quadratic) potential for phonons. It was demonstrated in Sec. \ref{sec:diat} how the anharmonicity in local vibrations can be responsible for thermal expansion. The thermal expansion coefficient $$\alpha
=\frac{1}{V}\frac{dV}{dT}$$ can be used in the dimensionless Gr\"{u}neisen parameter,
$$\Gamma = K\frac{\alpha}{C}$$ where the bulk modulus $B$ is defined  as the fractional change in volume $\delta V/V$  produced by a change $\delta p$ in pressure.
$$K=-\frac{\delta p}{\delta V/V}.$$

The Gr\"{u}neisen's law states that $\Gamma$ is temperature
independent. To show this we write the change in the
potential energy due to the deformation as
$$\delta U =K\left(\frac{\delta V}{V}\right)^2-
(d)(k_BT)\frac{(kT)^d}{(\hbar\omega
_0)^{d+1}}\frac{\partial\hbar\omega _0}{\partial V}\delta V$$
where $(d)$ is the system dimensionality. Here the first and second terms represent respectively the change in energy of a deformed isotropic body \cite{landauelast} and the change in the phonon energy
$U$ from Eq. (\ref{eq:Upot}). The latter is attributed to the change in characteristic frequency $\omega _0$. Direct minimization
with respect to $\delta V/ V$ gives
$$\frac{\delta V}{V}=\frac{d}{2K}\left(\frac{T}{T_D}\right)^{d+1}\frac{\partial\hbar\omega _0}{\partial
V} V.$$ Estimating here $|\partial \hbar\omega _0/\partial V|\sim
\hbar\omega _0/V$ yields $$\left|\frac{\partial\delta
V}{\partial T}\right|\sim \frac{1}{K}C$$ where $C$ is the
specific heat from Eq. (\ref{eq:DebyeT3}). Finally we arrive at
$|\Gamma| \sim 1$ in agreement with the majority of data for
different solids and with a more naive estimate in Sec. \ref{sec:diat}.

Note that in the above derivation, the effect of anharmonicity
was implicitly taken into account by assuming the limiting
frequency $\omega _0$ dependent on deformation. Note also that
while $|\alpha |\sim 1$, the sign of $\alpha$ remains
arbitrary. Indeed, it varies between different solids.

\subsection{Vibrations on fractals. Fractons}

\begin{figure}[tb]
\includegraphics[width=0.42\textwidth]{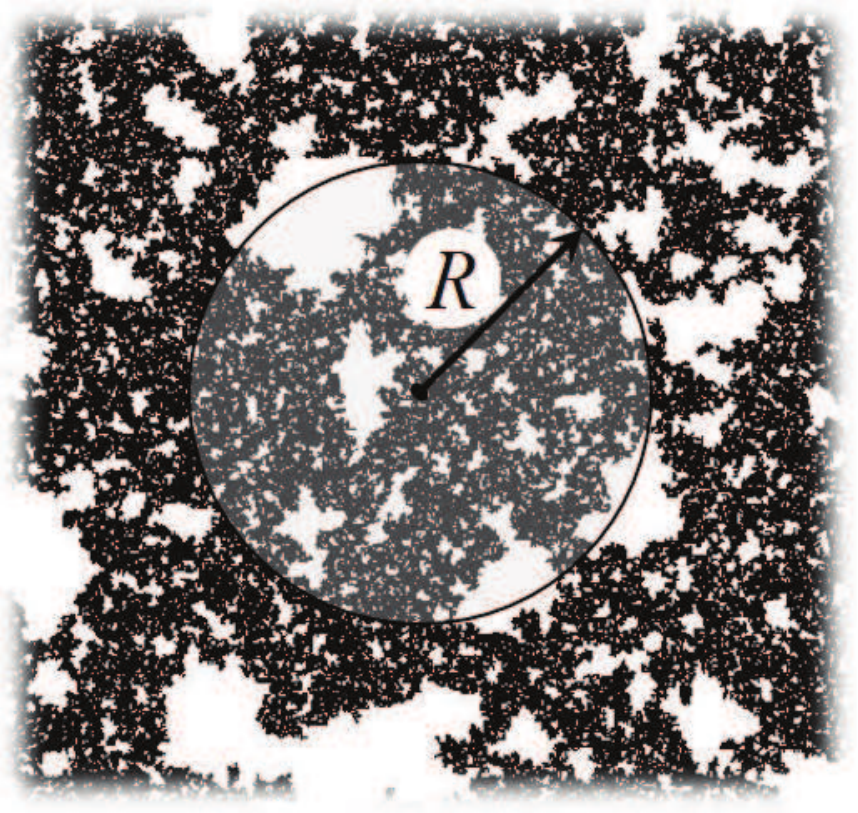}
\caption{A fractal structure representing polymers or composite materials. A semitransparent circle of radius $R$ encompasses the diffusion range over a certain time or that of vibration of certain frequency.
\label{Fig:fractal}}\end{figure}

Fractal networks can describe a variety of random (but not
necessarily only random) polymer-like systems illustrated in Fig.
\ref{Fig:fractal}. For such objects, the mass confined in a
volume of linear dimension $R$
is proportional to
$R^{\overline{d}}$ where $\overline{d}$ is called the principal
(or Hausdorff)\cite{mandelbrot} fractal dimension.
The architecture of
tree branches can give a simple example of the latter mass
scaling when the linear size $R$ is shorter than the linear
size of a tree.

As a first intuitive assessment, it is obvious that elastic waves on fractals will be
considerably different from that in a continuous medium. Indeed,
a fractal contains many pieces of one-dimensional architecture,
for which DOS $\propto \omega ^{d-1}=const$, hence, the
specific heat $C\propto T$. Because these pieces are connected
in the higher dimensionality objects, one can expect a
composite DOS $\propto \omega ^{d-1}$ with $d$ in between 1 and
3.

It is not immediately clear though, how to describe elastic waves in
fractals. In particular, the simplest (isotropic) form of the elastic
wave equation
\begin{equation}
\rho\ddot{u}=K\nabla ^2u\label{eq:elast}\end{equation} where
$u$ is the displacement and $\rho$ is the mass density,
contains the Laplacian $\nabla ^2$, which is the sum of second
derivatives. It is hard to imagine how to define the second
(or even the first) derivative for the case of an irregular
structure.

The following consideration shows how to treat the Laplacian on an intuitive basis (although the
question of derivatives in fractals has been addressed by pure
mathematics). The approach below draws a plausible analogy with the
diffusion equation that is also based on the Laplacian,
\begin{equation}
\dot{n}=D\nabla ^2n\label{eq:elast1}\end{equation} where $n$ is
the concentration and $D$ is the diffusion coefficient. This
diffusion problem in a fractal known also as the ``ant in the
labyrinth" has been solved by numerical modeling for many
different fractals. The main result of it is that the diffusion
is slower than that of regular lattices, which is natural due
to the presence of dead ends, closed loops, etc.

The latter
slowness is described by the diffusion coefficient
\begin{equation}D\propto R^{-\theta}\label{eq:rtheta}\end{equation}
where $\theta$ is the diffusion fractal index. We also note that the Eq. (\ref{eq:rtheta}) can be written as
$$\frac{\partial \eta}{\partial (Dt)}=\nabla^2 \eta.$$
Approximating $\partial n/\partial (Dt)\sim n /Dt$ and $\nabla^2\sim1/R^2$,
leads to the standard relation $Dt\sim R^2$. As a result, the
characteristic distance $R$ for propagation of a packet varies
with time as
\begin{equation}\label{eq:ant}
R\propto \sqrt{Dt}\propto t^{1/(2+\theta)}.
\end{equation}

We now note the
difference in the time derivatives in the vibrational
($\ddot{u}$) and diffusion ($\dot{u}$) equations. The time
Fourier transforms will then generate the frequency powers
$\omega ^2$ and $\omega $ in these equations respectively.
Replacing $t\sim\omega ^{-1}$ in Eq. (\ref{eq:ant})
and then replacing $\omega$ with $\omega ^2$ (to switch from diffusion
to vibrations) gives
\begin{equation}\label{eq:vib}
R\propto\omega ^{-2/(2+\theta)}.
\end{equation}

This gives the relationship between the linear dimension of a
vibrational state and its frequency. To normalize this quantity
to the number of vibrations per atom, we note that the number of
states per atom is proportional to $R^{-\overline{d}}\propto
\omega ^{2\overline{d}/(2+\theta)}$. Finally DOS is given by
\begin{equation}
D(\omega )\propto\frac{d}{d\omega}\omega
^{2\overline{d}/(2+\theta)}\propto \omega
^{\overline{\overline{d}}-1}\label{eq:fracDOS}\end{equation}
where
$$  \overline{\overline{d}}=\frac{2\overline{d}}{2+\theta}$$
is called the fracton dimension. According to Alexander and Orbach \cite{alexander},
$\overline{\overline{d}}\approx 4/3$ for a wide class of different fractal
structures (in spite of the fact that $\theta$ and
$\overline{d}$ vary considerably between different fractals in that
class). This kind of DOS and related heat capacity were indeed
observed in polymers. We note as well that the diffusion on fractals and fracton dimensionality describe the alternating current conduction on percolating clusters. \cite{schroder2008}

 \section{Electronic properties}\label{sec:eprop}
 \subsection{Band structure}\label{sec:bands}

\begin{figure}[bt]
\includegraphics[width=0.47\textwidth]{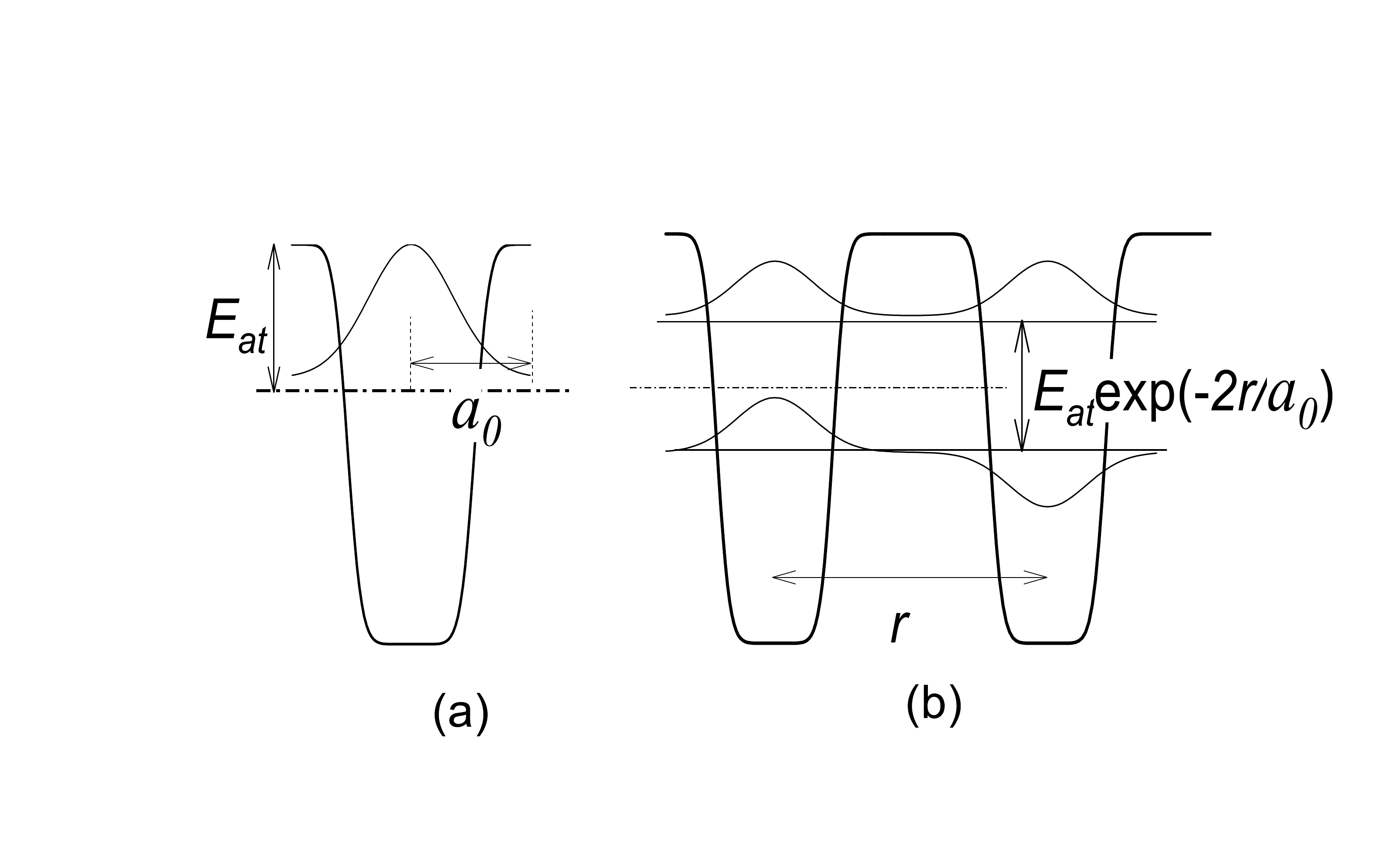}
\caption{Electron energy splitting in a system of two identical potential wells separated by distance $r$. Curved lines represent the wave functions. $a_0$ is the decay length of a wave function for a single atom.
\label{Fig:a0}}\end{figure}
The energy levels of electrons in crystals form finite width continuous bands separated by empty (forbidden) gaps. That energy structure can be described based on two different premises: (a) local basis approach where the bands are interpreted as the significantly broadened energy levels of the material's constituting atoms with the broadening caused by interatomic interactions, and (b) extended basis approach starting with the continuum of extended electron states fully collectivized between the constituting atoms, with boundaries caused by interference effects in a periodic atomic structure. We consider both the approaches.

\subsubsection{Local basis}\label{sec:locbas}
\begin{figure}[hbt]
\includegraphics[width=0.49\textwidth]{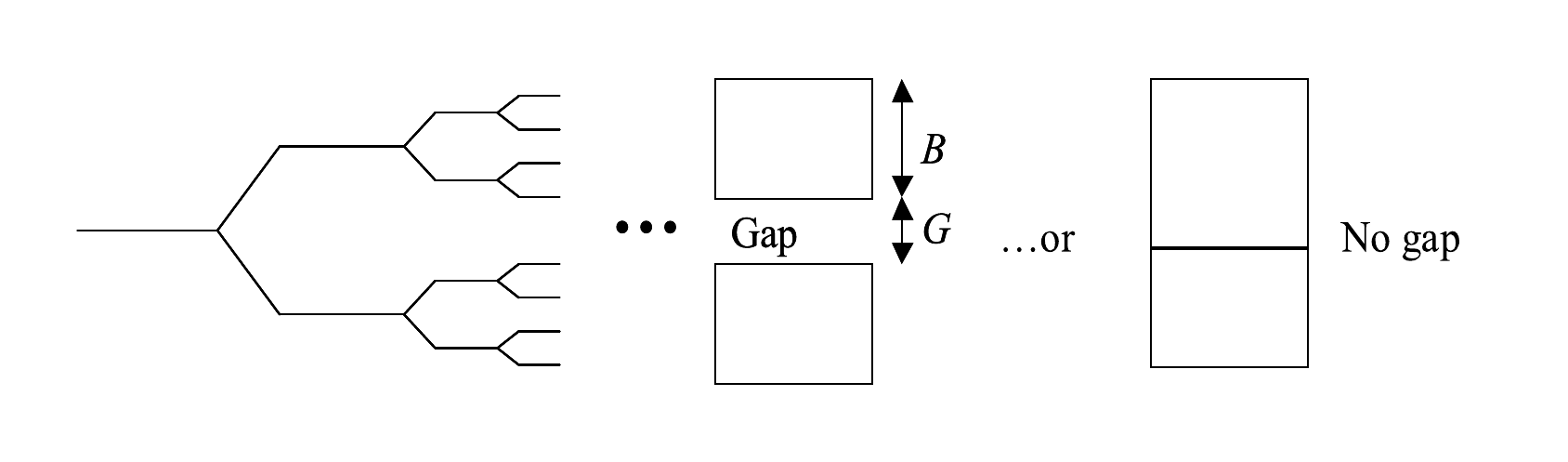}
\caption{Band formation as a consequence of the pairing energy
splitting . \label{Fig:band}}\end{figure}

As a thought experiment, we can form a material by first combining its constituting atoms in pairs, then forming pairs of pairs, pairs of pairs of pairs,  etc. For specificity, we assume atoms with
ionization energy $E_{at}\sim 10$ eV and electron radius $a_0\sim 1$ \AA (Fig. \ref{Fig:a0}).
As our first step we note that when two identical atoms form diatomic molecules their electronic states combine to form the bonding and antibonding orbitals separated by the energy gap
\begin{equation}G=E_{at}\exp(-2r/a_0)\label{eq:gap}\end{equation}
For the typical interatomic distances $r\sim 2$
\AA, their energy spectrum consists of a pair of energy levels
separated by the gap $G\sim 1$ eV.
Combining two diatomic molecules will introduce another splitting,
and so on, as shown in the diagram of Fig. \ref{Fig:band}. This
can result in either two bands separated by the gap $\sim G$, or in the overlapping bands
with no gap in between. Alternatively, the bands could
originate from the different atomic levels. This may change
certain band properties, but typically will not affect the
order of magnitude estimates here.

Within the band, the electron has the quasi-continuous spectrum with delocalized wave functions
characterized by the momentum $p=\hbar k$. The wave number $k$ varies
between zero and the maximum value $k_{max}\sim 1/a$, where $a$ is the lattice parameter (interatomic distance) for the case under consideration. $k_{max}$ corresponds
to the shortest possible wavelength $\lambda _{min}\sim a$.

Close to the band edge ($E_0$) we can express the electron energy as
$$E=E_0+b_2k^2+b_3k^3+b_4k^4+...$$ where $k\ll k_{max}$. Retaining only the quadratic term and setting $E_0=0$ enables one to present
$$b_2\sim \frac{\hbar ^2}{m^*}$$ where $m^*$ has the dimension
of mass and is called the effective mass. Because the only available
quantity of the right dimension is the electron mass, one can expect
$$m^*\sim m.$$ While generally true, the latter estimate can miss significant numerical factors, so effective masses can be quite
different, say $m^*\sim (0.1-0.3)\cdot m$. In some cases $m^*$ can be
significantly larger than $m$.

There is a simple approximate relationship between $m^*$ and the band
width $B$. Namely, if we extend the quadratic approximation
$$E-E_0=h^2k^2/2m^*$$ throughout the entire band by setting
$k\sim k_{max}\sim 1/a$, then
\begin{equation}\label{eq:efmas}B\sim \frac{\hbar ^2}{m^*a^2}.\end{equation}
The latter relationship, however approximate, adequately reflects the
trend: narrow band materials have large effective masses, while wide band
materials have light $m^*$.\cite{sze}

We shall end this subsection by noting that the energy gap $G=E_{at}\exp(-2r/a_0)$ will change with the system's deformations increasing or decreasing with interatomic distances $r$. That mechanism of deformation-induced variations in electronic energies is often called deformation potential. As such, it was present already in the earlier Fig. \ref{Fig:diatomic} where the increasing (right to the minimum) part of the energy can be associated with the bonding orbital energy, $E_{at}\exp(-2r/a_0)$. The deformation potential concept will be extensively used in Sec. \ref{sec:e-ph}.

\subsubsection{Extended basis}\label{sec:pert}

In the extended basis model, the electron states extend over all the atoms in a material. In the first approximation, they are described as plain waves $\psi = \exp(i{\bf kr})$ characterized by certain wave vectors ${\bf k}$ and their corresponding energies $E(k)=\hbar ^2k^2/2m$. These  plain waves are affected  by the presence of the lattice. We consider the lattice potential as a perturbation.

\begin{figure}[bt]
\includegraphics[width=0.47\textwidth]{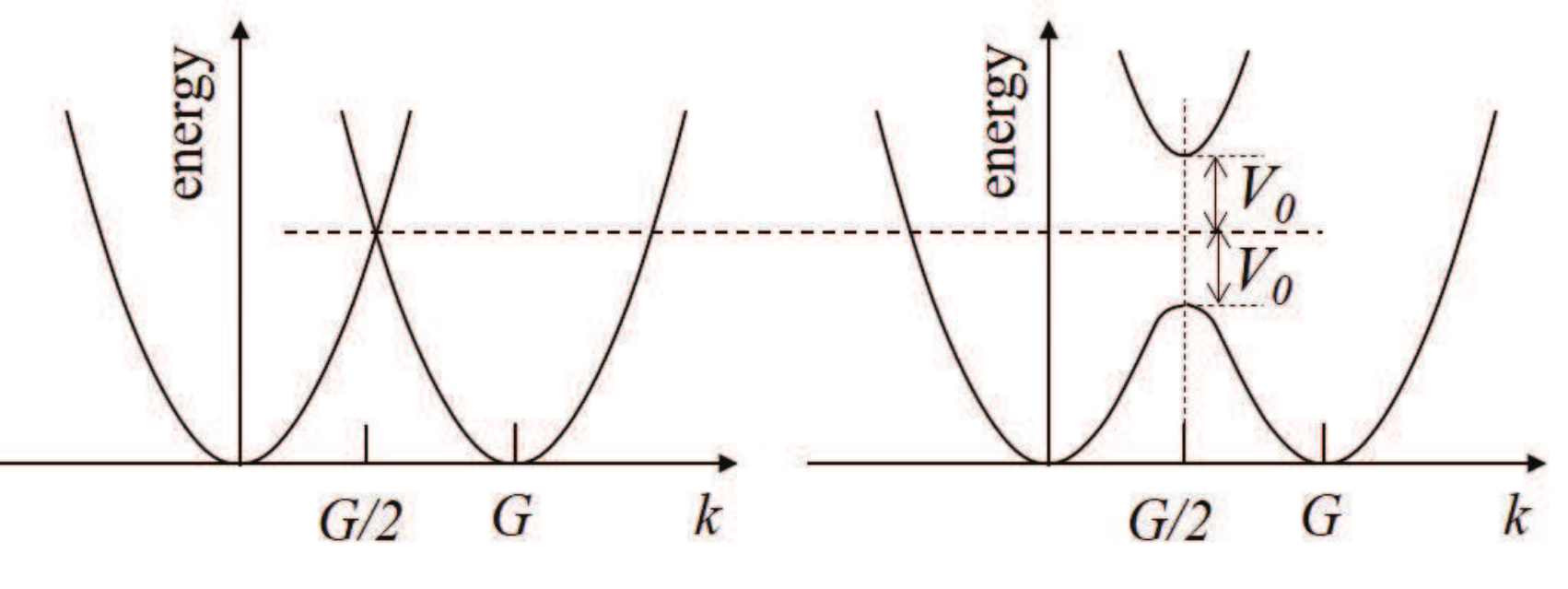}
\caption{Forbidden zone formation due to the repulsion of
degenerate electronic terms. \label{Fig:pert}}\end{figure}

When the lattice is present, the wavefunction should not change
under the translation ${\bf r}\rightarrow {\bf r+a}$ where ${\bf a}$ is the vector of
lattice translation. To reflect that translational invariance, the wave vector should change by \cite{kittel}
$\delta {\bf k}={\bf g}$, where $\bf g$ is called the vector of the reciprocal lattice; its absolute value is
$(2\pi /a)$. Therefore the description of electron energies will not change if we translate the original
dispersion parabola $E(k)=\hbar ^2k^2/2m$ by ${\bf g}$ as shown in Fig. \ref{Fig:pert}.
%At this point we
%employ the empty lattice concept, which, while not present physically, imposes
%certain translational invariance. (The empty lattice approximation considers the lattice potential to be periodic and weak. Its physical presence will be taken into account below.)

Fig. \ref{Fig:pert} (left) shows the dispersion curves shifted relative to each other by the wave vector ${\bf g}$. The degeneracy of the states belonging to these dispersion curves at the point of their intersection, ${\bf k=g/2}$ can be eliminated by small perturbations physically produced by a lattice. The corresponding perturbation theory for degenerate states \cite{landauqm,migdal} predicts
that the two curves will repel forming a gap $G$ in the energy spectrum.

More specifically the degenerate perturbation theory is described by Eq. (\ref{eq:tls}) in the appendix \ref{sec:app}. Denoting the unperturbed energies by
\begin{equation}\label{eq:dispdelta}E_{(g/2)\pm\delta}=\frac{\hbar ^2}{2m}\left(\frac{\pi}{a}\pm \delta\right)^2\end{equation} close to the intersection point ${\bf k}={\bf g}/2$, the
perturbed energies take the form
\begin{equation}
E_{\pm}(\delta )=\pm \sqrt{\left(\frac{E_{(g/2)+\delta}^0-E_{(g/2)-\delta}^0}{2}\right)^2+V_0^2}
\label{eq:deg}\end{equation}
where $\delta$ is a wavenumber measured from the intersection point.

The gap in the spectrum is $2V_0$. Furthermore, close to the intersection point, $\delta\ll g/2$, one can reduce the square root in Eq. (\ref{eq:deg}) to the form
\begin{equation}\label{eq:quad}\pm \left(V_0+\frac{4\pi ^2\hbar ^4 \delta ^2}{m^2a^2V_0}\right)\equiv \pm \left(V_0+\frac{\hbar ^2\delta ^2}{2m^*}\right)\end{equation} where the effective mass is given by
$$m^*=m\frac{V_0}{4\pi ^2\hbar ^2/ma^2}\sim m\frac{G}{B}.$$ Note that $m^*$
again has a negative correlation with the band width, which in
this case appears as $B=8\pi ^2\hbar ^2 /ma^2$.

As a general
note, it is due to the dimensionless factor of $G/B$ that the
ratio $m^*/m$ cannot be estimated with more certainty: any
dependence of the type $m^*=mf(G/B)$ where $f$ is a
`reasonable' (but arbitrary) function remains a possibility.

Also, we note that approximating the quadratic dispersion curve of Eq. (\ref{eq:quad}) over the entire allowed region $0<\delta <\pi /a$ results in the estimate for the allowed energy band $B=\pi ^2\hbar ^2/(m^*a^2)$, which is consistent with the alternative approach estimate in Eq. (\ref{eq:efmas}).

\subsection{Electron-phonon interaction}\label{sec:e-ph}

We have so far considered conduction electrons that move through a lattice whose atoms are frozen in place. In reality, a moving electron does interact with the ion cores; in an oversimplified picture, the electron pulls nearby positive ions and pushes negative ions away. Even if there are no ``real" phonons present, the electron will still interact with zero-point lattice vibrations.
According to the previous section (particularly, Eq.
(\ref{eq:gap})), the energy $E$ of the electron at the bottom
of a conduction band (or the hole at the top of the valence
band) will depend on the interatomic distance $r$. If we assume that the interaction is rather weak, and only retain the linear term  in the expansion of the exponential in equation (\ref{eq:gap}), then we can write
$\delta E\sim B\delta r/a$. Introducing here the dimensionless
dilation $u\equiv \delta r/a$, gives
\begin{equation}
\delta E= Qu.\label{eq:Q}\end{equation} The proportionality
coefficient $Q$ between the electron energy and the dilation is
known as the deformation potential. Note that the latter linearization could be equally applied to the rising part of potential energy in Fig. \ref{Fig:diatomic}. Therefore all the consideration addressing electron-phonon interactions below will qualitatively apply not only to solids but to molecules as well, showing again their representativeness of condensed matter physics.

A more realistic
description would distinguish between different types of
deformation (and introduce the corresponding deformation
potential tensor $Q_{ij}$ as a matrix of coefficients between the
$i$th component of deformation and the energy of the electron
moving along the $j$th axis). Neglecting this anisotropy we roughly
estimate a scalar $Q\sim B$, that is $Q$ of the order of
several eV.

The above deformation effect is referred to as the
electron-phonon interaction. Indeed, each phonon can be
considered as a source of deformation that can change the
electron energy. Along these lines, any arbitrary deformation
can be represented as a superposition of partial deformations
related to a (complete) set of phonons. In other words, the deformation potential,
is a Hamiltonian of the electron-phonon interaction that can be used to
describe the electron-phonon scattering.

\subsubsection{Polaron effect}
\begin{figure}[bt]
\includegraphics[width=0.47\textwidth]{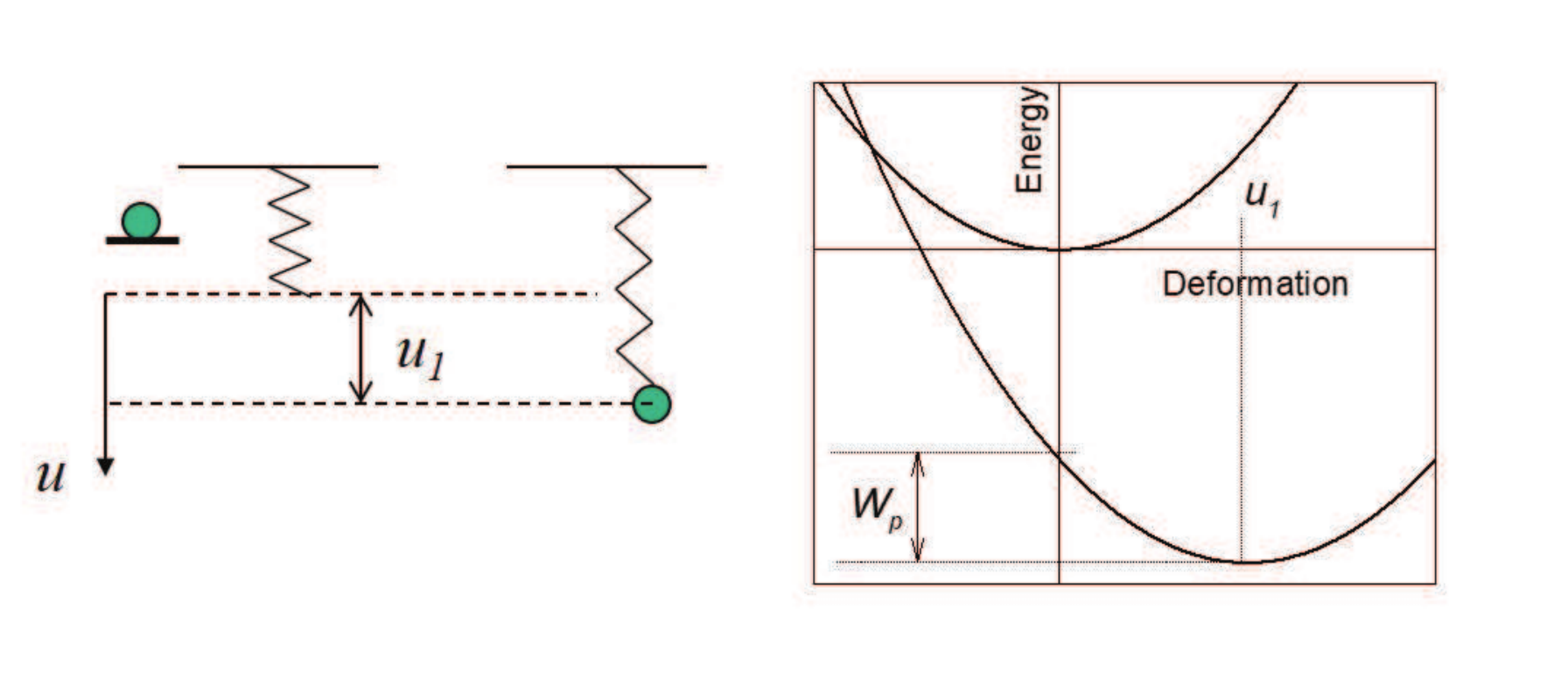}
\caption{Left: Deformation of a spring with a hanging particle
as a model for the electron-lattice interaction. Right:
Energy vs. deformation for the case when the electron is not
present (upper parabola), and in the presence of the electron.
\label{Fig:pshift}}\end{figure}
One consequence of the deformation potential
interaction is the polaron effect.\cite{landaupekar} We will start with noting that the
 linear deformation interaction $Qu$, takes place in
parallel with the quadratic increases in elastic energy,
$\kappa u^2/2$, as stated by Hook's law, where $\kappa$ is the elastic
modulus. Therefore the total energy of our system (electron and lattice),
\begin{equation}
W=E_0-Qu+\frac{\kappa u^2}{2}\label{eq:springel}\end{equation}
becomes formally equivalent to that of a particle of mass $m$
hanging on a spring in a uniform gravitational field. In this latter case
$Q=mg$ and $u$ is the spring elongation. This analogy extends
rather far. Similar to the equilibrium spring elongation, one
can find the stationary ($dW/du=0$) material deformation $u_1$ that
minimizes the system energy:
\begin{equation}
u_1=\frac{Q}{\kappa}\quad {\rm and}\quad
W_{min}=E_0-\frac{Q^2}{2k}\equiv E_0-W_p
\label{eq:pshift}\end{equation} where $W_p$ is called the
polaron shift (Fig. \ref{Fig:pshift}). Here, $E_0$ is the
electron energy in the system prior to the deformation. For
example, $E_0$ is the energy of the electron at the bottom of
the conduction band, or it can be the energy of the electron
localized at a defect state (located inside the energy gap) before the deformation occurred.
 Note that while the system
energy changes by $W_p$, the electron energy change is twice as
large, \begin{equation}
E_0-Qu_1=E_0-2W_p.\label{eq:ee}\end{equation} From this
consideration we conclude that it may be energetically
favorable for the system to create a local deformation thus
decreasing the electron energy. However, the parameters $Q$ and
$\kappa$ in the above consideration are not strictly defined; for
example, they appear unrelated to a particular type of the
electron wave function, localized or delocalized, number of
electrons involved, their kinetic energies, spin states, etc.

\subsubsection{Negative-$U$ centers}
To be more specific, we will next consider the electron states
localized at some defects. In that case, the electron wave
function has a finite size, typically of the order of the
corresponding de Broglie wavelength, $R\sim \hbar
/\sqrt{m^*E}\sim 10 \AA$ where $E\lesssim G$ is the defect
binding energy, $G\sim 1 - 2 $ eV is the forbidden gap, and
$m^*\sim (0.1 - 0.5)m$ is the effective mass. We assume that
$Q\sim 1-2$ eV remains a good estimate for the deformation
potential, and will use $k\sim 3-5$ eV for the elastic modulus
(same order of magnitude as the ionization energy). The corresponding polaron
shift is $$W_p\sim 0.1-1\quad {\rm eV}.$$ With these figures in
mind, we consider the possibility of the two electrons pairing
at the same defect center. This phenomenon is called the
negative Hubbard energy (or negative correlation energy, or
negative $U$)\cite{hubbard,anderson} and does take place in a variety of materials,
specifically in chalcogenide glasses, some of A$^2$B$^6$
semiconductors, and possibly in superconducting ceramics. Its
essence is that instead of one electron occupying each defect center, the electrons prefer to form pairs (of opposite
spins) filling one half of such centers while leaving the other
half empty; the singly occupied electron states appear
energetically unfavorable.

\begin{figure}[bt]
\includegraphics[width=0.32\textwidth]{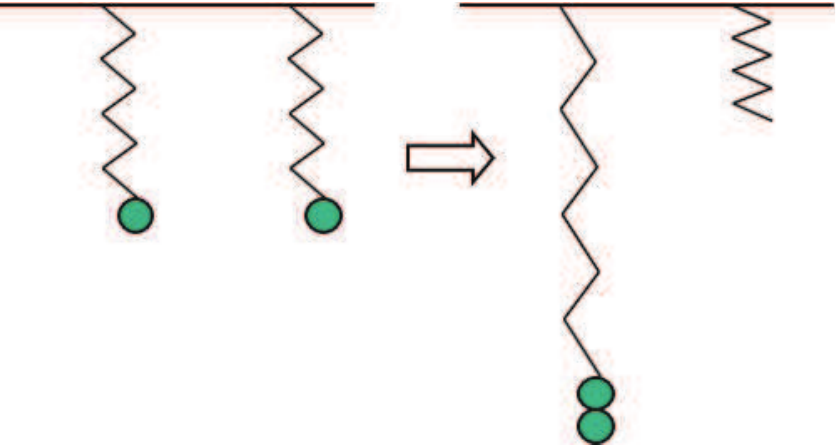}
\caption{Mechanism of electron pairing represented in terms
of elastic springs. \label{Fig:pairs}}\end{figure}

The pairing mechanism is illustrated in Fig. \ref{Fig:pairs},
starting with the analogy of two point masses hanging on
elastic springs. Using classical mechanics, we can see that the energy of the two individual masses, each hanging on one spring is higher than the energy of the two masses hanging together on the same spring. In our case at hand, the pairing energy gain can be estimated based
on the approach similar to that in Eq. (\ref{eq:springel}),
\begin{equation}
W_n=nE_0-nQu+\frac{\kappa u^2}{2}+U_c\delta _{n,2}.
\label{eq:springel2}\end{equation} Here $n=0,1,2$ is the center
(spring) occupation number and $U_c$ is the Coulomb repulsion
that takes place when $n=2$. In the simplest approximation,
$$U_c\sim\frac{e^2}{\varepsilon R}\sim 0.2 \quad {\rm  eV}$$ assuming the
dielectric permittivity $\varepsilon \sim 10$.
 Optimizing
Eq. (\ref{eq:springel2}) yields the equilibrium energies,
\begin{equation}
W_{n,min}=nE_0-n^2W_p+U_c\delta
_{n,2}.\label{eq:shifts}\end{equation} The correlation energy
is defined as the difference between the state of a pair of
electrons and that of the two separate electrons,
\begin{equation}
U=W_{2,min}-2W_{1,min}=-2W_p+U_c.\label{eq:U}\end{equation}
With the above estimates for $W_p$ and $U_c$ in mind, both the
cases of $U<0$ and $U>0$ become possible, depending on
particular values of the material parameters. Softer materials
with small $\epsilon$ (glasses, polymers) and locally soft
configurations (locally soft atomic potentials due to special
defect structure configuration or random softening in
disordered materials such as chalcogenide glasses) are
especially likely candidates for the negative-$U$ phenomenon.
The latter entails numerical consequences, such as the
diamagnetism coexisting with a finite electron density at the
Fermi level, abnormally large difference between the
characteristic energies of absorption and photoluminescence
(PL), and some others.

\subsubsection{Electron transition energies}\label{sec:etren}

\begin{figure}[bt]
\includegraphics[width=0.5\textwidth]{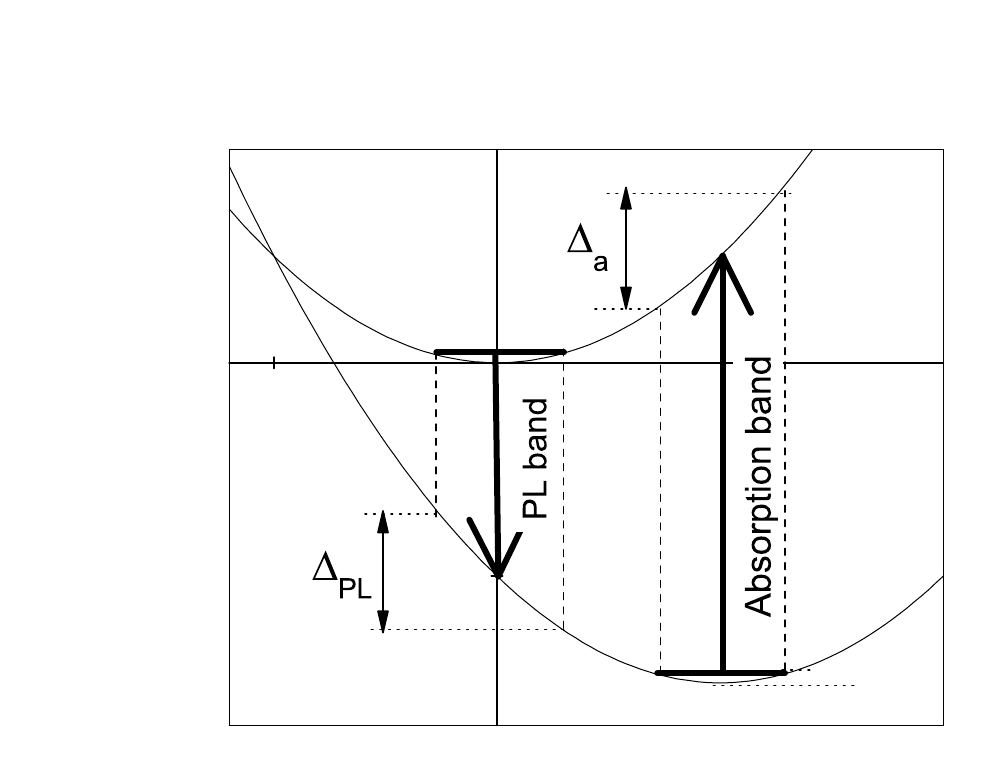}
\caption{Absorption and PL transitions associated with defect
state in a deformable medium. \label{Fig:stokes}}\end{figure}

In the case of a considerable polaron effect, the same electron
state is characterized by considerably different energies
corresponding to its (1) thermal ionization ($E_T$), (2)
photo-ionization ($E_{PI}$) and (3) PL ($E_{PL}$). The nature
of these differences is illustrated in Fig. \ref{Fig:stokes}.

\begin{figure}[bt]
\includegraphics[width=0.5\textwidth]{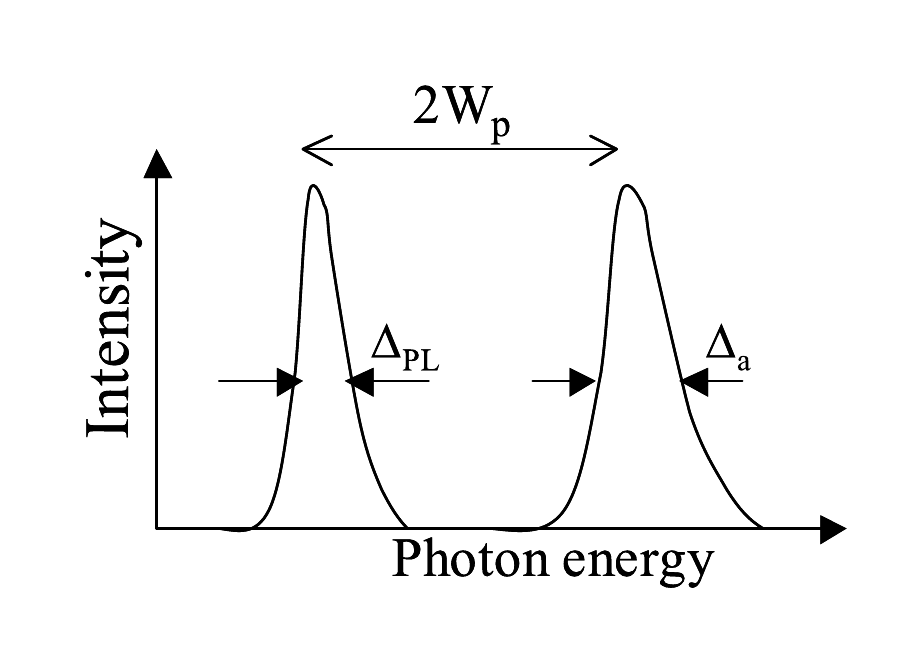}
\caption{Absorption and PL bands corresponding to the
transitions shown in Fig. \ref{Fig:stokes}.
\label{Fig:bands}}\end{figure}

The absorption process is depicted by an upward arrow from the
state of the most likely (equilibrium) deformation of the
bottom parabola. If we assume that the deformation does not change, the u coordinate will remain unchanged in the course of the transition, and the arrow from the bottom to the top parabola will be almost vertical. This is consistent with the standard adiabatic
approximation: the electron transitions are so fast that the
atomic nuclei do not have time to follow. (The optical transition time corresponding to a $600$ nm wavelength is of the order of $10^{-15}$ s). Similarly, the PL
transition is depicted by the downward arrow; its energy is
noticeably lower than that of the absorption. The
thermal ionization energy is, by definition, the minimum work we need to
perform in order to take the electron from the center. (This
minimality does not rule out the non-adiabatic transitions.) As
a result, $E_T$ is given by the energy difference between the
two parabola minima. Photo-ionization on the other hand, is the photo-induced decrease of the charge of the system by one electron. In our case, the upper parabola corresponds to the energy of the system without one electron, while the bottom parabola corresponds to the energy of the system with the electron. To calculate the length of the photo-ionization arrow, we have to first traverse the energy $E_T$ (from the bottom minimum to the top minimum) and them add the energy $1/2\kappa u_1^2$ that will bring us to the top parabola. Since $u_1=Q/\kappa$ and $W_p=Q^2/2\kappa$, we can see that
$$E_{PI}=E_T + \frac{1}{2}u_1^2= E_T+W_p$$
From Fig. (\ref{Fig:stokes}) we see that $E_T=E_{PL}+W_0$, therefore
$$E_{PL}+2W_p.$$ The polaron shift can
be estimated as a half difference between the centers of the
absorption and PL bands as shown in (Fig. \ref{Fig:bands}).

To estimate the PL and absorption bands, we note that the
system is not strictly localized at the parabolic minimum but it is rather exercising small vibrations with characteristic amplitude $\lambda$. For low temperatures, ($k_BT\ll \hbar\omega$) $\lambda$ can be estimated using the zero-point energy of the harmonic oscillator $E=\hbar \omega$. From the De Broglie wavelength $\lambda=\hbar/p$ we can obtain:
$$\lambda =\sqrt{\frac{\hbar}{M\omega}}$$
For high temperatures ($k_BT\gg
\hbar\omega$), we can estimate $\lambda$ using the energy of the harmonic oscillator: $k_BT\sim 1/2 k \lambda^2$ which gives us:
$$ \lambda\sim\sqrt{\frac{2k_BT}{\kappa}}$$
In both cases $M$ is the
characteristic atomic mass and $\omega$ is the characteristic
frequency of atomic vibrations. %By linearizing the parabolic
%dependence in the proximity of $u=Q/k$, the corresponding width
%of the absorption and PL bands becomes

We will now estimate the width of the absorption and PL bands, which according to figure (\ref{Fig:stokes}) are approximately equal for small deformations. Considering a small change in the energy of the harmonic oscillator near the point $u_1$ gives us
$$\Delta _{PL}\sim \Delta _a\sim \frac{1}{2}u_1^2-\frac{1}{2}(u_1-\lambda)^2=\kappa \lambda u_1=\lambda Q=\frac{\lambda}{u_1}W_p$$
%frac{\lambda}{Q/\kappa}2W_p=\frac{\lambda}{u_1}2W_p.$$
We note that generally $\lambda /u_1$ is much less than
unity, because $u_1$ can be comparable to the atomic length,
while the atomic vibration amplitude $\lambda$ is considerably
smaller, say $\lambda /u_1\sim 0.1 - 0.3$.

\subsubsection{Multi-phonon electron transitions}
Consider the probability of the electron downward transition between two
energy levels separated by the energy gap $\Delta E\gg
\hbar\omega$. It is proportional to the probability $p_N=p_1^N$
of emitting
$$N_{ph}=\frac{\Delta E}{\hbar\omega}$$
phonons needed to dissipate the energy. The latter probability
can be equally represented in the form
\begin{equation}
p_N=\exp \left(-\gamma\frac{\Delta E}{\hbar\omega}\right),\quad
\gamma\equiv
\ln\left(\frac{1}{p_1}\right)\label{eq:mph}\end{equation}
common to many other adiabatic electron transition
probabilities (such as e. g. multi-photon ionization of atoms).

\begin{figure}[bt]
\includegraphics[width=0.27\textwidth]{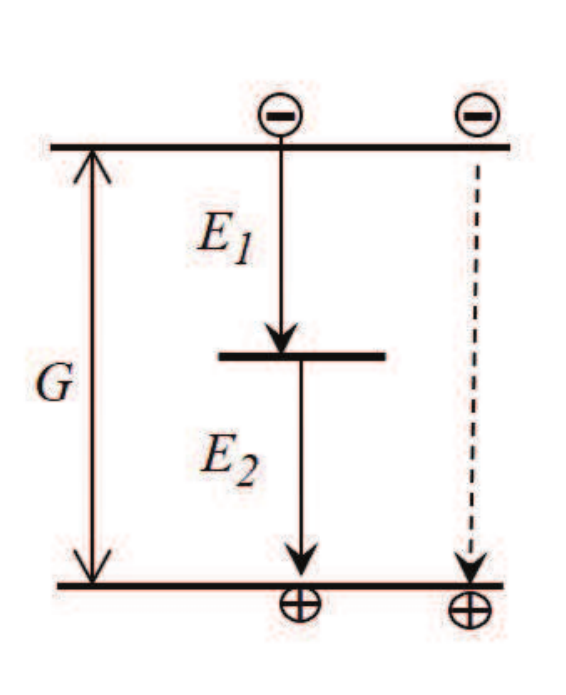}
\caption{Electron-hole recombination through a midgap defect level vs the direct, without defect recombination with exponentially lower probability $\sim \exp(-\gamma G/\hbar\omega )$
deformation. \label{Fig:recomb}}\end{figure}

In the physics of semiconductors, one of the most important
applications of the energy dependence in Eq. (\ref{eq:mph}) is
that the energy levels in the middle of the forbidden gap
($\Delta E\sim G/2$) appear the most efficient recombination
centers as is generally accepted on empirical grounds. Eq.
(\ref{eq:mph}) does predict that the energy $G/2$ optimizes
between the probabilities of an electron going downward to the defect level $\Delta E_1\sim G/2$)
and then further downward ($\Delta E_2\sim G-\Delta E_1\sim G/2$) to recombine with a hole in a valence band.
It is illustrated in Fig. \ref{Fig:recomb}

\begin{figure}[bt]
\includegraphics[width=0.42\textwidth]{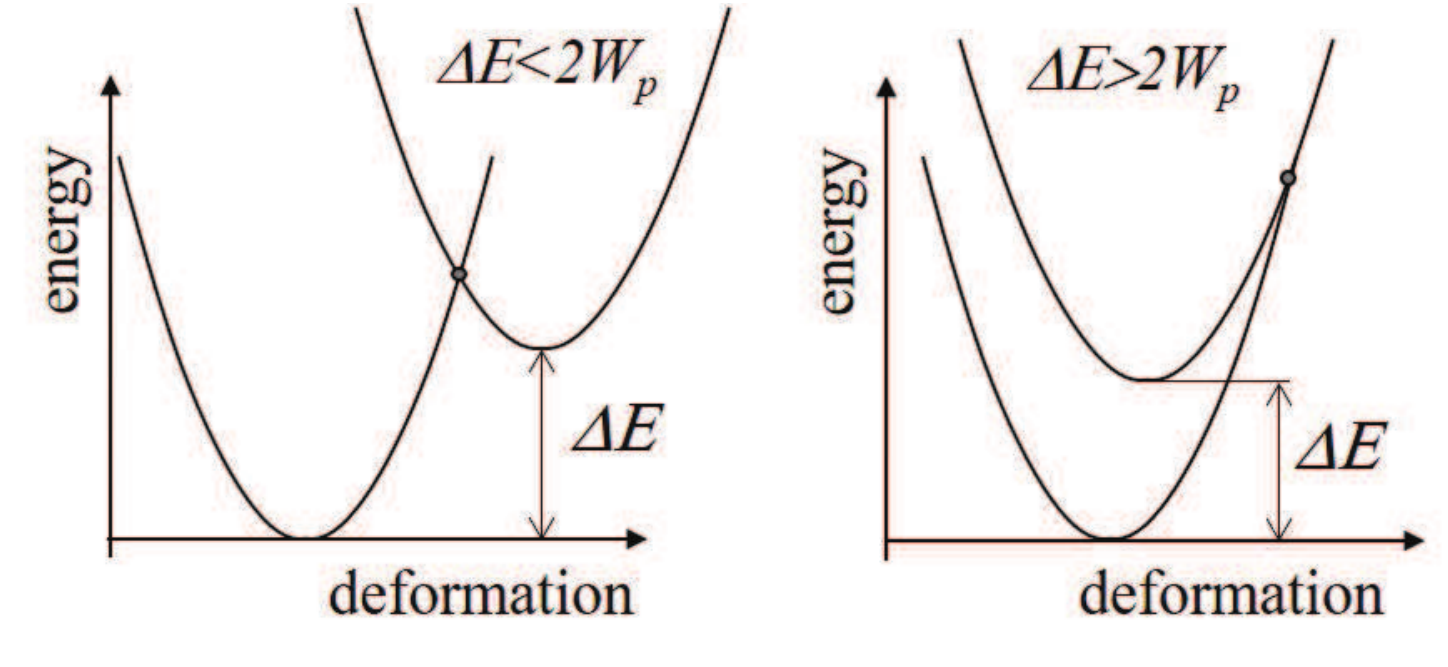}
\caption{Electron transitions for the cases of $\Delta E> 2W_p$
and  $\Delta E< 2W_p$ shown in the electron energy level
diagram and the corresponding electron terms vs. system
deformation. \label{Fig:mph}}\end{figure}

The above reasoning leading to Eq. (\ref{eq:mph}) can be
somewhat deceptive, since in a deformable medium the required
change in the electron energy can be significantly different
from that of the system energy due to the electron-lattice interaction.
As is illustrated in Fig. \ref{Fig:mph} the transition
probability can be more accurately related to overcoming the potential barrier formed by a point of intersection between the initial and final electron states in a deformable medium. \cite{mott,baran1986} Omitting the details, that barrier can either decrease or increase with $\Delta E$,
depending on the relationship of $\Delta E$ and the energy
$2W_p$ representing the difference between the energy of bare
(before deformation) and equilibrium {\it electron} energy. In
particular, when $2W_p>\Delta E$, the corresponding transition probability
$$p\propto \exp\left(-\gamma\frac{2W_p-\Delta
E}{\hbar\omega}\right)\propto\exp\left(\gamma\frac{\Delta
E}{\hbar\omega}\right)$$ {\it increases} with $\Delta E$. On
the contrary, for the case of $2W_p<\Delta E$ the original dependence in Eq. (\ref{eq:mph}) takes place.
In terms of parameters, both cases are possible and can
give rise to a number of observed phenomena.
%({\it Discuss the Meyer - Neldel rule}).

\subsubsection{Small polaron collapse}

Consider a polaron in the ideal crystal.\cite{chat2018} Its energy is given by
the expression that generalized the above analysis to include
the electron wave function $\psi ({\bf r})$,
\begin{equation}
W=\frac{\hbar ^2}{2m}\int (\nabla\psi )^2d^3r-Q\int u\psi
^2d^3r +\frac{k}{2}\int u^2d^3r.\label{eq:polen}\end{equation}
Here the first term describes the electron kinetic energy, the
second one stands for the deformation potential interaction,
and the last term accounts for the elastic energy. To minimize
its total energy the system adjusts the deformation shape
$u({\bf r})$ to a certain equilibrium configuration determined
by the condition that the variation of $W$ with respect to $u$
is zero,
$$-Q\psi ^2+ku=0\quad {\rm i. e.}\quad u=\frac{Q}{k}\psi ^2.$$
Substituting this back into Eq. (\ref{eq:polen}) yields the
system energy
\begin{equation}
W=\frac{\hbar ^2}{2m^*}\int (\nabla\psi
)^2d^3r-\frac{Q^2}{2k}\int \psi
^4d^3r\label{eq:polen1}\end{equation} that has to be further
minimized with respect to the wave function $\psi$.

Instead of reducing the latter functional to a differential
equation, we consider approximating it by using a one-parameter trial
wave function. The parameter is the localization radius $R$, so
that
$$\psi ^2 \sim \frac{1}{R^3}$$ in accordance with the
normalization condition. Because $(\nabla \psi )^2\sim \psi
^2/R^2$ and the integration reduces to just multiplication by
the volume $R^3$ (in which the wave function is finite), the
entire functional is estimated as
\begin{equation} W\sim\frac{\hbar
^2}{m^*R^2}-\frac{Q^2}{kR^3}.\label{eq:polen3}\end{equation}

\begin{figure}[bt]
\includegraphics[width=0.32\textwidth]{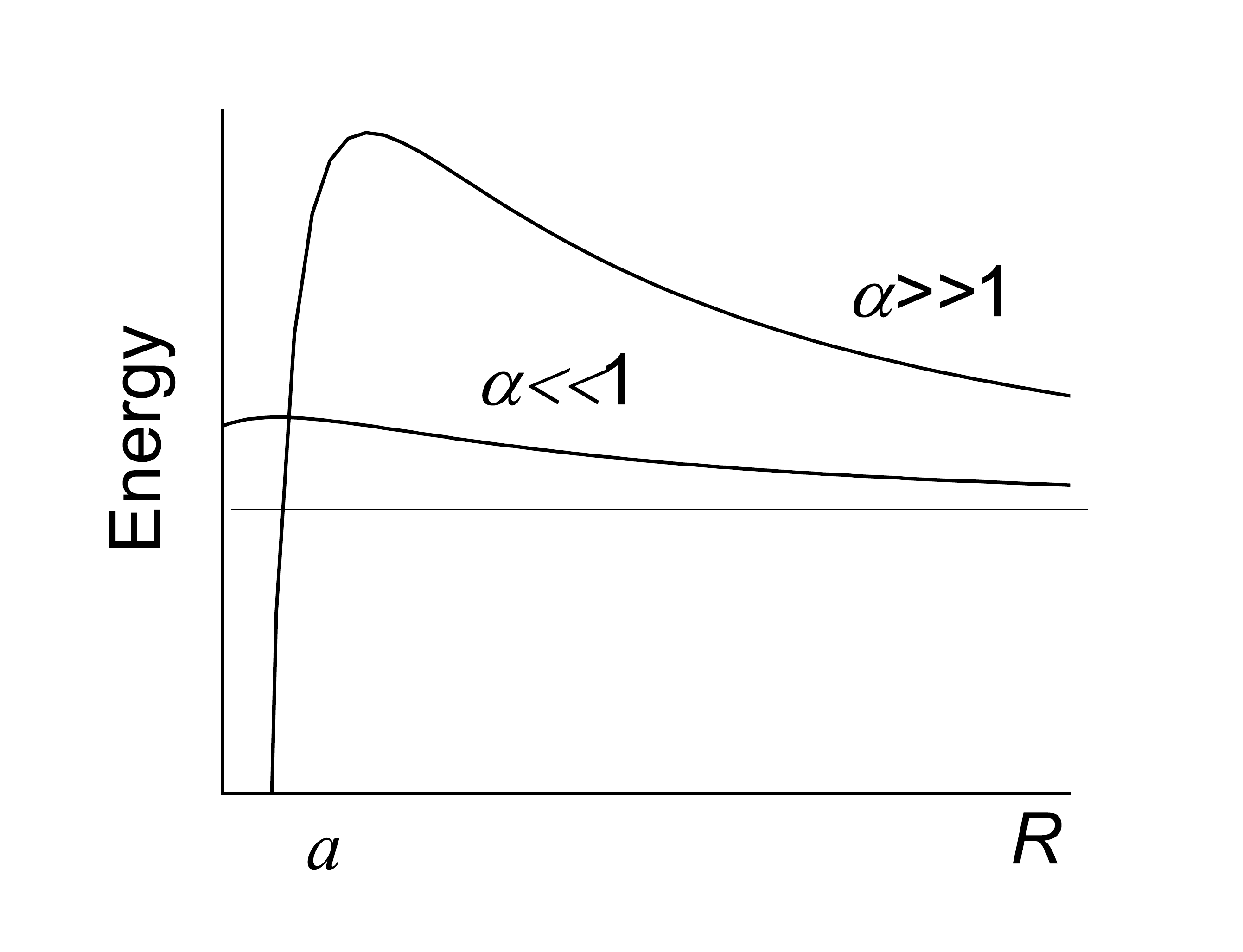}
\caption{The electron-lattice system energy as function of the electron
localization radius for the cases of $\alpha \gg 1$  (strong coupling) and $\alpha \ll 1$ (weak coupling).
\label{Fig:col}}\end{figure}

Formally, Eq. (\ref{eq:polen3}) predicts the system collapse to
zero radius ($R\rightarrow 0$) that provides the energy
minimum. This polaron collapse is however restricted from
bellow by the discretness of atomic lattice, $R\sim a$. For
$R\lesssim a$, further collapse of the electron wave function
does not decrease the system energy. The system behavior is
then decided by the relative contribution of the two terms in
Eq. (\ref{eq:polen3}) at $R\sim a$, that is by the dimensionless
coupling parameter
\begin{equation}
\alpha = \frac{Q^2/k}{\hbar ^2/m^*a^2}\sim
\frac{W_p}{B}\label{eq:Fr}\end{equation} where $B$ is the band
width. As illustrated in Fig. \ref{Fig:col}, when $\alpha \gg
1$, the system spontaneously creates a small ($R\sim a$) polaron
state of strongly deformed lattice, decreasing the electron
energy by $W_p$ and confining its wave function to the
dimension of the crystal elemental cell. In the opposite case
of $\alpha \ll 1$, the system does not create any deformation
about the electron. (In fact, the states with $\alpha\lesssim
1$ can still have some degree of accompanying deformation;
their analysis goes beyond the present frameworks).

The criterion of $\alpha > 1$ can be given a simple
interpretation. It takes the energy of $\sim B\sim \hbar
^2/ma^2$ to localize the electron within a region of linear
dimension $a$. Such a localized state will become energetically
stable if the subsequent polaron gain $W_p$ overbalances the
energy loss $B$, i. e. $\alpha >1$.

\subsubsection{Polarons in polar media}

The above polaron consideration was based on the concept of
electrically neutral deformation typical of covalent materials.
In ionic crystals, such as NaCl and likewise, there can be
another type of deformation leading to a considerable electric
polarization. Namely, if the electron (hole) is confined to
some local region, positive (negative) ions will be attracted
towards its electric charge distribution, while the opposite
ions will be pushed away from it. This creates the electric
polarization whose field will result in  a self-consistent
potential well for the original electron (hole) supporting its
localization.

In other words, such a dipole type deformation can be called
the optical polaron as related to the deformation induced
polarization characteristic of optical phonons. The previously
described case of electrically inactive deformation can be
referred to as the acoustic polaron, since this deformation can
be represented as a superposition of acoustic phonons.

To describe the optical polaron we note that in response to the
electric potential $\phi =e^2/r$ a material develops the
counter potential $$\phi
'=\left(\frac{1}{\epsilon}-1\right)\frac{e^2}{r},$$ such that the
sum  $\phi +\phi '$ gives the well-known screened potential
$e^2/\epsilon r$ where $\epsilon$ is the dielectric permittivity.
Of the polarization potential $\phi '$, only the inertial part
that is not accompanying the electron dynamics can contribute
to a static potential well (the fast part of polarization will
follow the electron motion, maybe partially contributing to its
effective mass). To account for the slow polarization component
we will subtract from $\phi '$ its part related to the
dielectric permittivity at very high frequency, $\epsilon
_{\infty}$. After such a subtraction, the stationary part of
the polarization potential becomes
\begin{equation}
\phi '=\left(\frac{1}{\epsilon _0}-\frac{1}{\epsilon
_{\infty}}\right)\frac{e^2}{r}\equiv -\beta\frac{e^2}{r}
\label{eq:polpot}\end{equation} where $\epsilon _0$ is the static
dielectric permittivity. In ionic crystals, the coupling
parameter $\beta $ is not small, because the difference between
$\epsilon _0$ and $\epsilon _{\infty}$ is quite significant.
The heavy ions significantly contribute to the static
polarization, but are rather immaterial in the high-frequency
response (unlike the typical covalent crystals where both the
above components are mostly due to the electrons and are very
close to each other).

With Eq. (\ref{eq:polpot}) in mind, the total polaron energy can be
written as
\begin{equation}
W=\frac{\hbar ^2}{2m^*}\int (\nabla\psi )^2d^3r-\beta e^2\int
\frac{\psi ^2({\bf r})\psi ^2({\bf r'})}{|{\bf r}-{\bf
r'}|}d^3r'd^3r.\label{eq:polpol}\end{equation} We have omitted
here the part of electrostatic energy related to the material
polarization  ($\int P^2d^3r/8\pi$ where $P$ is the vector of
electric polarization), which is proportional to the parameter
$\beta ^2$ and thus is relatively small.

Using the order-of-magnitude estimate $\psi ^2\sim R^{-3}$ and
replacing integration with the multiplication by the electron
localization volume $R^3$ yields
\begin{equation}
W\sim\frac{\hbar ^2}{m^*R^2}-\frac{\beta e^2}{R}.
\label{eq:polpol1}\end{equation} The latter equation is
formally the same as that of the hydrogen atom with the nucleus
charge $\beta e$. Based on this analogy, the polaron energy
becomes
$$E\sim \frac{\beta ^2e^4m^*}{\hbar ^2}.$$

\subsection{Disordered Systems}

\subsubsection{Hopping conduction}

\begin{figure}[bt]
\includegraphics[width=0.27\textwidth]{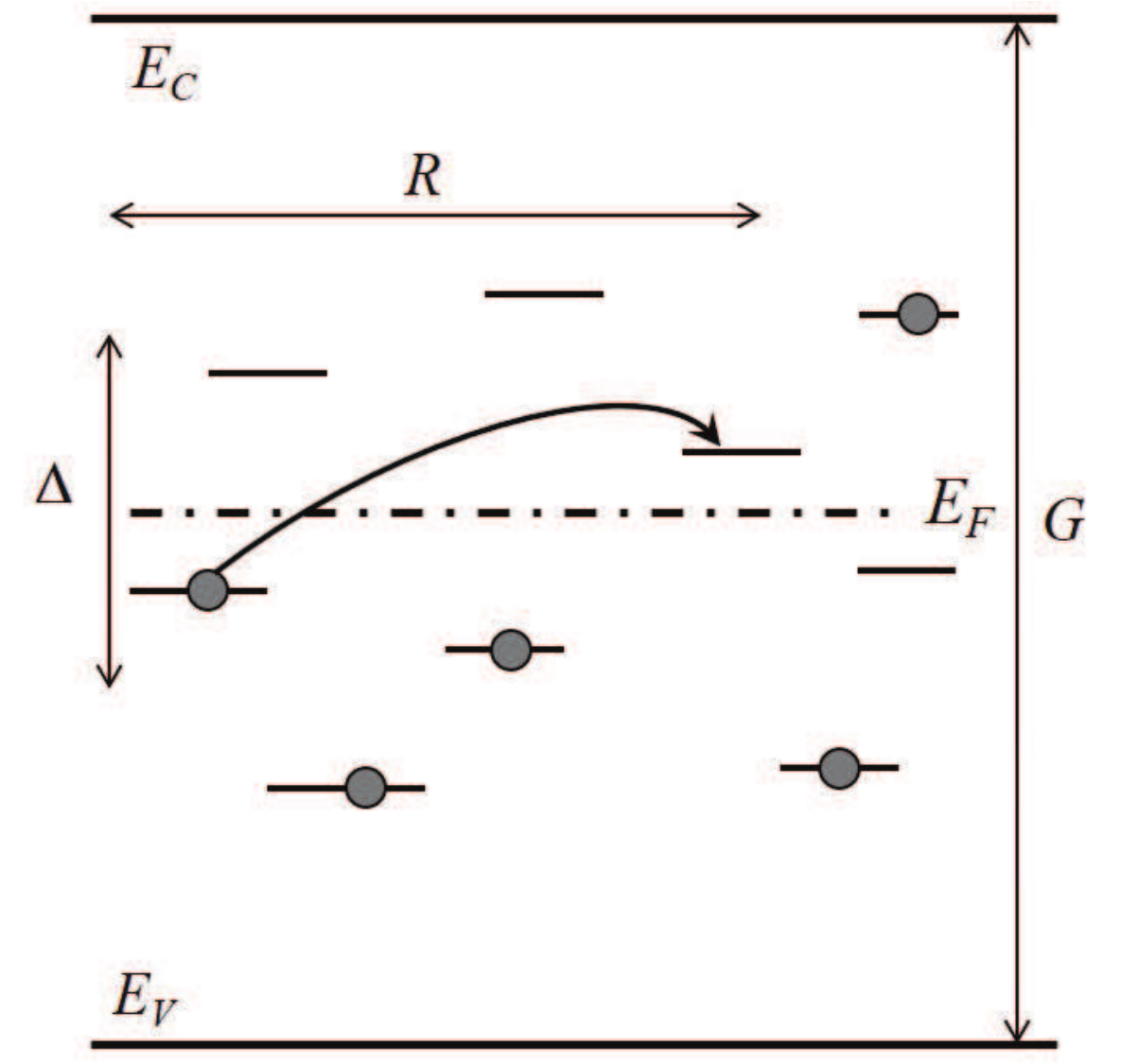}
\caption{Electron hopping between the localized states in the
proximity of Fermi level. \label{Fig:hop}}\end{figure}

Localized states in many noncrystalline semiconductors form
quasi-continuous spectra of energy levels in the proximity of Fermi
energy. When the energy gaps between the Fermi level and
the mobility edges are too large to make the thermal activation of
electrons and holes efficient, the electron hopping between
localized states becomes a dominant transport mechanism. Its
description utilizes two parameters, which are the width
$\Delta$ of the energy band within which the hopping takes
place, and the characteristic hopping distance $R$ (Fig.
\ref{Fig:hop}).

The approach is based on estimating the electron hopping
diffusion coefficient
$$D\sim \nu R^2$$
where $\nu$ is the hopping frequency. The electron mobility can
be then expressed through the Einstein relation
$$\mu=\frac{D}{k_BT}$$ so that the conductivity $\sigma =
ne\mu$, with the effective electron concentration $n$ corresponding
to the states in the energy band $\Delta$, i. e. $n\sim
g\Delta$ where $g$ (cm$^{-3}$eV$^{-1}$) is the electron density
of states. We assume $g$ to be energy independent.

The probability of inter-center transition is a product of
probabilities of tunneling ($\exp(-2r/a)$)  and thermal
activation ($\exp(-\Delta /k_BT)$) where $a$ is the radius of
the electron wave function on a defect. The latter implies that
the electron trajectory composed of sequential inter-center
transitions can be represented by a set of resistors in series
with the hardest link corresponding to the activation by the
entire bandwidth $\Delta$. As a result the probability of
hopping becomes
\begin{equation}
\nu =\nu _0exp\left(-\frac{\Delta}{k_BT}-\frac{2r}{a}\right)
\label{eq:hop}\end{equation} where $\nu _0$ is the frequency of
(hopping) attempt, of the order of the characteristic phonon
frequency 10$^{13}$ s$^{-1}$.

Given the probability in Eq. (\ref{eq:hop}), it is obviously
desirable to make both $\Delta$ and $R$ as small as possible.
However, it is rather unlikely that both of these quantities
are simultaneously small. To the contrary, it is quite
intuitive that the closest center (minimum $R$) will typically
have a significant energy difference with that of the original
electron location, and vice versa, to find a center that is
very close in energy, one will have to look far enough from its
original location, hence, large $R$.

The typical situation corresponds to the condition that the
center of destination is found with certainty in a volume of
$R^3$ within the energy band $\Delta$, that is
$$gR^3\Delta \sim 1.$$
The latter condition presents the competition between $\Delta$
and $R$ explicitly: the smaller the one the bigger the other.

Expressing $\Delta \sim 1/R^3g$ and substituting this into Eq.
(\ref{eq:hop}) gives the exponent in which one term increases
with $R$ while the other decreases. This competition resolves
in the optimum values given (to within numerical factors) by
\begin{equation}
\Delta =k_BT\left(\frac{T_0}{T}\right)^{1/4}, \quad
R=a\left(\frac{T_0}{T}\right)^{1/4}, \quad T_0\equiv
\frac{1}{ga^3k_B}.\label{eq:hopop}\end{equation} We note that
the effective temperature $T_0$ is very much higher than $T$.
Indeed, assuming the typical order-of -magnitude estimate
$g\sim 10^{19}$ cm$^{-3}$eV$^{-1}$ and $a\sim 10 \AA$ yields
$T_0\sim 10^{6}$ K. Hence, $(T_0/T)^{1/4}\sim 10$ when $T\sim
100K$: the electron hops across the distances and energies by
order of magnitude greater than its wave function length
and temperature respectively.

Substituting the optimum values from Eq. (\ref{eq:hopop}) into
Eq. (\ref{eq:hop}) leads to the famous Mott law \cite{mott} for the
variable range hopping conduction, $$\sigma=\sigma _0\exp\left[
- \left(\frac{T_0}{T}\right)^{1/4}\right],\quad\sigma _0\sim
ge^2\nu _0k_BT\left(\frac{T_0}{T}\right)^{1/4}$$ verified
experimentally for many different amorphous semiconductors.

\subsubsection{Band Tails}

\begin{figure}[bt]
\includegraphics[width=0.27\textwidth]{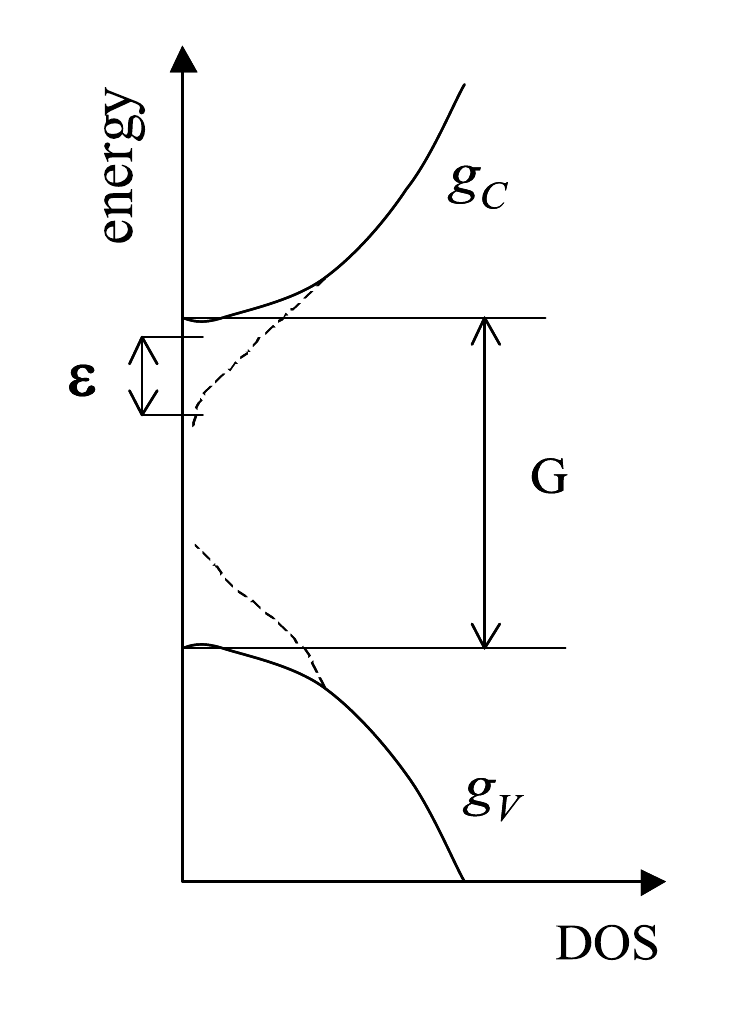}
\caption{Tails of density of states (DOS) of conduction and
valence bands (shown in dash) decay into mobility edge with the
characteristic energy scales $\varepsilon$.
\label{Fig:tail}}\end{figure}

In disordered systems, such as doped semiconductors (impurity
atoms occupying random positions) or amorphous silicon, or
window and other glasses, the concept of forbidden gap undergoes
certain changes compared to a perfectly ordered ideal crystal.
This happens for obvious reasons: the system ceases to be
invariant with respect to translations, imperfections are
everywhere in the form of valence angle fluctuations, dangling
bonds, fluctuations in chemical composition, etc. \cite{mieghem}In
particular, well defined edges of conduction and valence
bands become fuzzy, and instead of sharp boundaries give rise
to rather diffuse tails of states gradually decaying into what
is now called the mobility gap (Fig. \ref{Fig:tail}).

\begin{figure}[bt]
\includegraphics[width=0.32\textwidth]{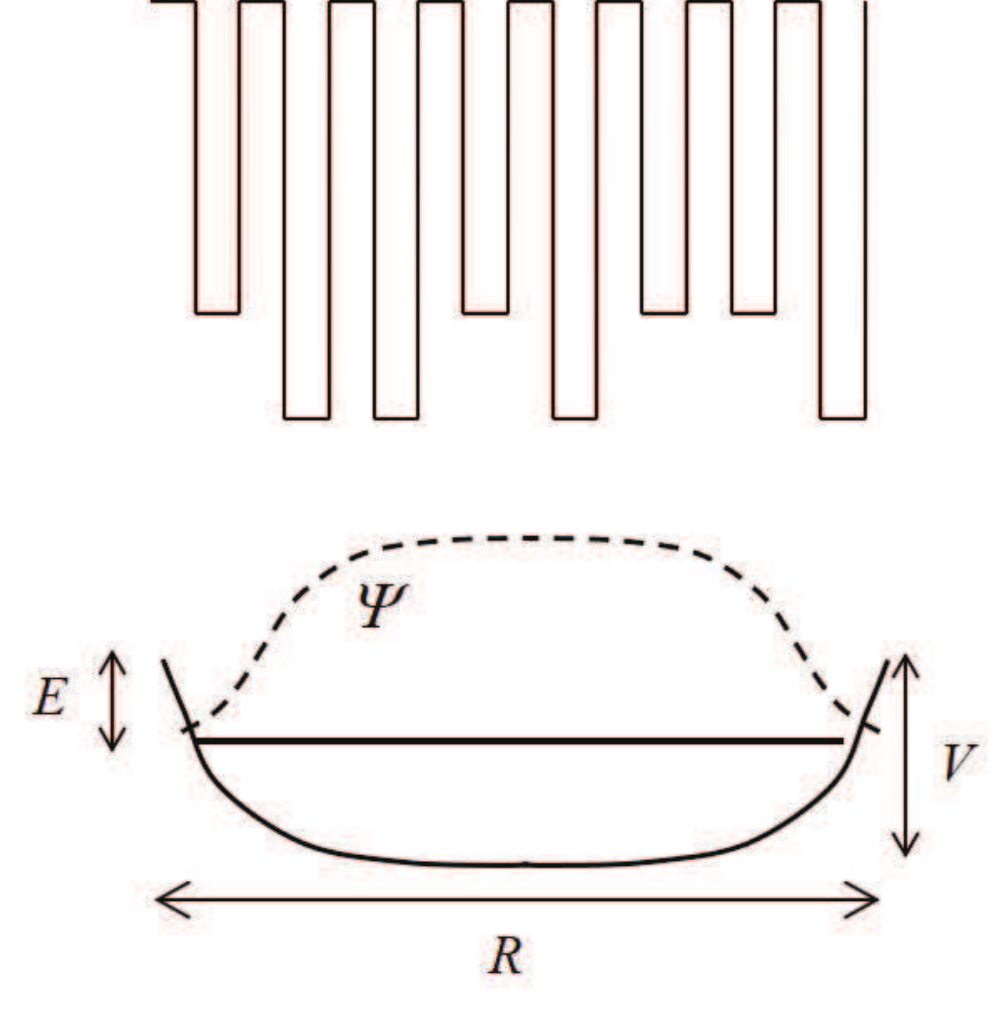}
\caption{Top: Fluctuation in impurity atom concentration.
Bottom: Impurity atoms have stronger short-range potential
wells that altogether act as the effective potential well of
depth $V$ and length $R$ capable of localizing the electron,
whose wave function is shown by dashed line.
\label{Fig:tailwell}}\end{figure}

%There exists a number of different flavor theories of band
%tails.
 A number of different theories, from different perspectives, are in existence.
 We describe one such more intuitive theory, according to
which band tails originate from local fluctuations in atomic
potentials felt by the electron or hole. These potentials are
thought of as short-range potential wells (say, associated with
impurity atoms), whose concentration $n$ undergoes statistical
fluctuations with the characteristic dispersion $\sqrt{N}$,
where $N$ is the total number of atoms in a volume $R^3$ (Fig.
\ref{Fig:tailwell}).

To describe DOS in the band tail we proceed from the probability of
Gaussian fluctuations
\begin{equation}p\sim \exp\left[-\frac{(\delta
N)^2}{\overline{N}}\right]=\exp\left[-\frac{\delta
nR^3}{\overline{n}}\right]\label{eq:Gaussfl}\end{equation} for
the fluctuation of $\delta N$ particles against the background
of their average number $\overline{N}$ in the volume $R^3$. The greater the fluctuation $\delta n$ in defect
concentration, the deeper the effective potential well for the
electron and correspondingly the deeper the energy level in the
tail.

\begin{figure*}[htb]
\includegraphics[width=0.62\textwidth]{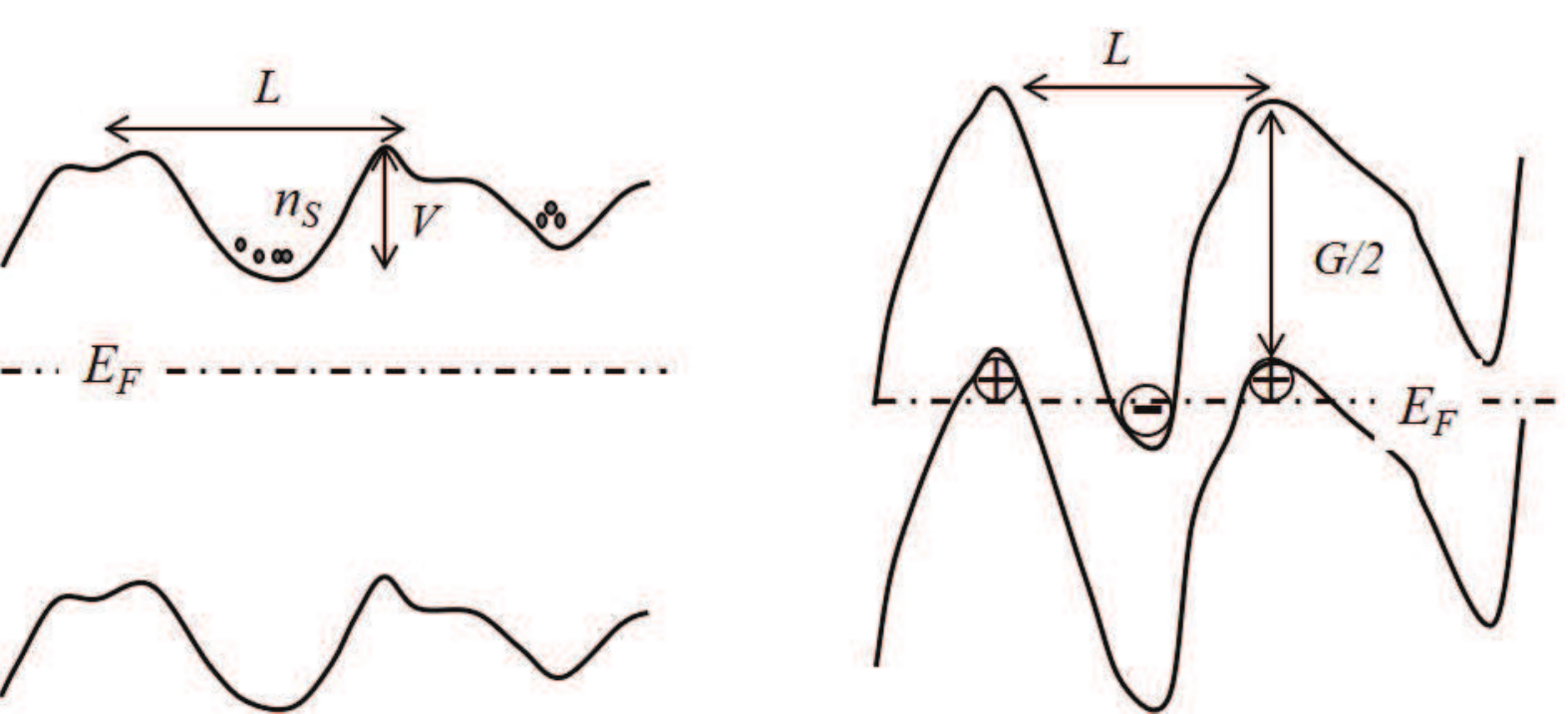}
\caption{Left: Long range random potential fluctuations in the
case of significant screening charge carrier concentration
$n_S\gg n_G$ (but still $n_S\ll N$). Right: Same in the case of
very low screening concentration $n_S\ll n_G$.
\label{Fig:potrel}}\end{figure*}

The latter probability exponentially increases when
$\delta n$ or $R$ decrease. However, they cannot be
made arbitrarily small because the corresponding potential well
then become too shallow (small $\delta n$) and too narrow
(small $R$) to accommodate a state of given energy
\begin{equation}
E\sim \frac{\hbar ^2}{m^*R^2}-\gamma \delta n.
\label{eq:tailen}\end{equation} Here, the first and second terms
represent the kinetic and potential electron energy
respectively, and $\gamma$ is the proportionality coefficient
between the potential well depth $V$ and the concentration
fluctuation $n$: $V=\gamma \delta n$.

Expressing $\delta n$ from Eq. (\ref{eq:tailen}) and
substituting into Eq. (\ref{eq:Gaussfl}) gives
$$p\sim \exp\left[-\frac{(E-\hbar
^2/mR^2)^2R^3}{\overline{n}}\right].$$ Optimizing the latter
with respect to $R$ gives $R\sim \hbar /\sqrt{m^*E}$ and the exponentially decaying DOS,
\begin{equation}
g(E)\propto \exp\left(-\sqrt{\frac{E}{E_0}}\right), \quad
E_0\equiv \frac{m^{*3}\gamma ^4\overline{n}^2}{\hbar
^6}.\label{eq:tailDOS}\end{equation}

Note that while the above outlined efficient optimization of the exponent can result in a rather approximate result missing a significant numerical multiplier, \cite{mieghem} which is practically important when the exponent is large. Also, the presence of such a multipler is rather unusual against the background of the majority of approximate  methods where such multipliers are typically of the order of unity.

\subsubsection{Random Potential}

Here, we briefly discuss variations in potential energy due to randomly distributed charged centers in semiconductors. \cite{shklovskii1984}
We assume that such centers are randomly distributed in doped and
compensated semiconductors. In a volume of linear dimension
$L$, their average number is
$\overline{N}=\overline{n}L^3$ where $n$ is their concentration, and the characteristic
fluctuations are given by: $\delta N =\sqrt{N}$.

The corresponding charge
fluctuation is a sum of positive and negative centers'
contributions. They are close to each other when the degree of compensation is high enough, $\delta
Q=q(\delta N_{+}+\delta N_{-})\approx 2q\sqrt{N}$. The corresponding electric potential fluctuations become,
\begin{equation}\delta \phi\sim \frac{\delta Q}{\epsilon L}\sim \frac{\sqrt{\overline{n}L}}{\epsilon }\label{eq:potfluc}\end{equation}
where $\epsilon$ is the dielectric permittivity. We observe that
$\delta \phi$ {\it diverges} with $L$.

The above potential fluctuations (Fig. \ref{Fig:potrel})
do not become infinitely large due to charge carrier screening that we now
discuss. Lets assume first that the compensation is not exact
and there exists a small concentration $n_S=|N_{+}-N_{-}|\ll
N_{+}$ of mobile charge carriers.  The screening is achieved when the
average charge concentration fluctuation $\delta N/L^3$ becomes comparable
with $n_S$, i. e.
\begin{equation}L\sim
\overline{n}^{-1/3}\left(\frac{\overline{n}}{n_S}\right)^{2/3}.\label{eq:len}\end{equation}
If the ratio $\overline{n}/n_S$ is not too large (compensation
not too strong) then $L\sim \overline{n}^{-1/3}$ is of the
order of the average inter-center distance, and the
corresponding $\phi$ is not particularly significant. However,
in the case of really strong compensation and large ratio
$\overline{n}/n_S\gg 1$, the screening length gets much longer,
and -using Eq. (\ref{eq:len})- the potential fluctuation becomes
\begin{equation}\delta\phi \sim \frac{q\overline{n}^{2/3}}{\epsilon
n_S^{1/3}}.\label{eq:potfl}\end{equation}

The latter estimate holds until the fluctuation amplitude
becomes comparable to $G/2$ where $G$ is the forbidden gap, i.
e. $$n_S\gg
n_G=\overline{n}\left(\frac{q\overline{n}^{1/3}}{\epsilon
G}\right)^3.$$ In the opposite limiting case of $n_S\ll n_G$, (where $n_G$ is the
electron concentration within the forbidden gap $G$)
the electric potential fluctuations of the conduction and
valence bands overlap forming metal type areas where the band
edges cross the Fermi level (Fig. \ref{Fig:potrel}).

In reality, it is rather hard, if possible at all, to achieve a
significant degree of compensation leading to strong enough
potential fluctuations in Eq. (\ref{eq:potfl}). However, some
materials, such as A2B6 semiconductors, are known for their
ability of `self-compensation'. Practically, this means that
they strongly resist doping: introducing donors in these
materials trigger the energetically favorable defect creation.
The newly created defects of acceptor type compensate the
original donors: the electrons leave positively charged donor
centers and localize at acceptor defects making them negatively charged
. The degree of self-compensation is so high that
introducing up to $N\sim 10^{19}$ cm$^{-3}$ of doping impurities
results in as low as 10$^{14}$ cm$^{-3}$ charge carrier
concentration identifiable with $n_S$; hence, $n_S\ll N$.

\section{Conclusions}
 We hope that the above material reflects to a certain extent the diversity of models, methods and techniques in condensed matter physics and shows how the inductive approach becomes particularly fruitful for such a subject. We fully expect to hear that the reader will find our selection biased and emphasizing some topics at the expense of others. We hope that other instructors and researchers will make broadly available their qualitative methods related to a variety of other subjects and that this effort will provide an example of how that can be achieved.

 The above set of examples appears accessible to college and grad students and can be blended indeed into
the existing condensed matter curricula. We have successfully used
those examples teaching upper undergrad and graduate level courses, such as Solid State Physics, Condensed Matter Physics, and Qualitative Methods in Physics. We
have observed higher student enthusiasm toward both in-class and home assignments compared to the standard
curriculum items. We attribute that positive feedback to an
appreciation for the shorter learning curve provided by
the qualitative methods along with their inductive approach being natural to our cognition. The injection of
qualitative methods will arm not only students, but acting professionals as well, with the toolkit needed as they
deal with the demands of technology -driven environments.

\section*{Acknowledgements}
We are grateful to Diana Shvydka and A. V. Subashiev for useful discussions.

\section*{Appendices}

\appendix
\section{Perturbation
theory for adjacent levels}\label{sec:app}
%\begin{figure}[bt]
%\includegraphics[width=0.4\textwidth]{Fig7.eps}
%\caption{Perturbation for adjacent levels.} \label{Fig:perturbation}\end{figure}
\begin{figure}[b!]
\includegraphics[width=0.4\textwidth]{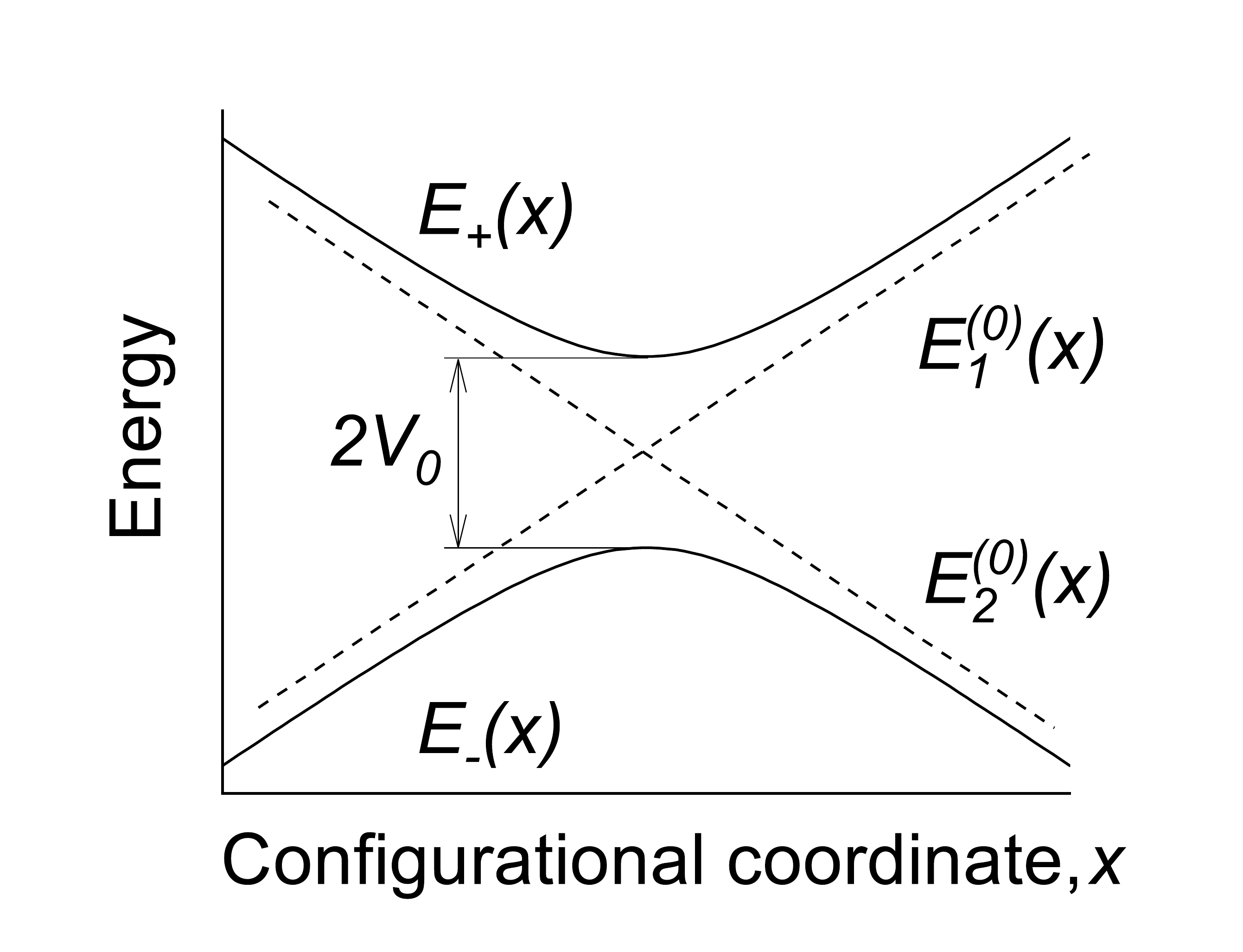}
\caption{Perturbation for adjacent levels.} \label{Fig:perturbation}\end{figure}

We start with $E_1(r)$ and $E_2(r)$, the energies of two states vs some parameter $r$, the physical meaning of which can vary between different problems. For example, $E_1(r)$ and $E_2(r)$ can stand for the energy levels of 2p and 1s states in a well formed by a Coulomb attractive potential with a short range hump of height $U_0$ at its center as a function of $U_0$, in which case these states can have the same energies at a certain $U_0$.  Alternatively, $E_1(r)$ and $E_2(r)$ can stand for the energies of two electron terms in a diatomic molecule vs. its interatomic distance as implied in Fig. \ref{Fig:band}; they can equally describe the energy levels in a double well potential given in Eq. (\ref{eq:DWPE}). Similarly, they can represent the electron energies $E_1(k)$ and $E_2(k)$ vs. their wave vectors $k_1$, $k_2$ as implied in Sec. \ref{sec:pert}.

In all such cases where two energy levels are very close to each other, one can neglect the effects of all other energy levels in a system dealing with a two-state basis, which makes the system solvable exactly. Because of its simplicity, the description in this Appendix turns out to be extremely useful with a variety of applications; yet it remains presented insufficiently in standard curricula.

Following Landau and Lifshitz, \cite{landauqm} we consider a point $r_0$ where $E_1(r)$ and $E_2(r)$ are close but not equal, such that a change $\delta r\ll r$ makes $E_1$ and $E_2$ equal. That common value of energy $E$, is the eigenvalue of the Schr\"{o}dinger equation
\begin{equation}(\hat{H_0}+\hat{V})\Psi = E\Psi .\label{eq:Sch}\end{equation} where $\hat{H_0}$ is the unperturbed Hamiltonian and $\hat{V}$ is a small perturbation caused by $\delta r$. The unperturbed wave functions satisfy
$$\hat{H_0}\Psi _{1,2}=E_{1,2}\Psi _{1,2}.$$ As a first-order approximation, we consider the superposition
$\Psi = C_1\Psi _{1}+ C_2\Psi _{2}.$ Substituting $\Psi$ into Eq. (\ref{eq:Sch} ) leads to
$$c_1(E_1+\hat{V}-E)\psi_1+c_2(E_2+\hat{V}-E)\psi_2=0.$$
Multiplying this equation by $\psi^*{_1}$ or $\psi^*{_1}$, and integrating (taking into account that $\psi_1$ and $\psi_2$ are orthogonal) one gets,
\begin{eqnarray}c_1(E_1+V_{11}-E)+c_2V_{12}=0 \\ c_1 V_{21}+c_2(E_2+V_{22}-E)=0 \label{c1c2}\end{eqnarray}
where $V_{ij}= \int\psi^*{_i}\hat{V}\psi_j d^3 r$. Since $\hat{V}$ is hermitian, $V_{11}$ and $V_{22}$ are real, and $V_{12}=V^*_{21}$. For the system of the above two equations to have a non-trivial solution, we need the discriminant to be equal to zero, which yields a quadratic equation for $E$ with two solutions,
\begin{eqnarray}E=E_{\pm}\equiv \frac{1}{2} (E_1 + E_2 + V_{11} + V_{22}) \nonumber \\
\pm \sqrt{\frac{1}{4}(E_1 - E_2 + V_{11} - V_{22})^2 + |V_{12}|^2}\end{eqnarray}

By denoting $E_1 + V_{11}= E_1^{(0)}$, $E_2+V_{22}=E_2^{(0)}$ and $V_{12}V^*_{21}=V_0^2$ finally yields,
\begin{equation}\label{eq:tls}E_{\pm}= \frac{E_1^{(0)} + E_2^{(0)} }{2}
\pm \sqrt{\frac{(E_1^{(0)} - E_2^{(0)})^2}{4} + V_0^2}.\end{equation}

The latter result is illustrated in Fig. \ref{Fig:perturbation}. To the accuracy of notations, it reduces to that in Eqs. (\ref{eq:DWPE}) and (\ref{eq:deg}).
%{(G/2)+\delta}^0$ _{(G/2)-\delta}^0$,

\end{document}